\documentclass{llncs}

\usepackage{amsmath}
\usepackage{latexsym}
\usepackage{theorem}
\usepackage{amssymb}
\usepackage{txfonts}
\usepackage{upgreek}
\usepackage{stmaryrd}
\usepackage{hyperref}
\usepackage{appendix}
\usepackage{tikz}
\usepackage{bussproofs}
\usepackage{aeguill}
\usetikzlibrary{arrows,positioning}

\newcommand{\comment}[1]{}

\newcommand{\eg}{{\em e.g.} }
\newcommand{\ie}{{\em i.e.} }
\newcommand{\etal}{{\em et al} }

\newcommand{\dom}{\mr{dom}}

\renewcommand{\a}{\rightarrow}
\newcommand{\A}{\Rightarrow}

\renewcommand{\AA}{\Leftrightarrow}
\newcommand{\la}{\leftarrow}

\renewcommand{\to}{\mapsto}

\newcommand{\I}[1]{[\![#1]\!]}

\newcommand{\bs}{\boldsymbol}

\newcommand{\ex}{\exists}
\newcommand{\all}{\forall}
\newcommand{\ou}{\vee}

\newcommand{\et}{\wedge}

\renewcommand{\th}{\vdash}

\newcommand{\sle}{\subseteq}

\newcommand{\tge}{\unrhd}

\newcommand{\tgt}{\rhd}

\newcommand{\al}{\alpha}
\renewcommand{\b}{\beta}
\newcommand{\g}{\gamma}
\newcommand{\G}{\Gamma}

\newcommand{\D}{\Delta}

\newcommand{\z}{\zeta}
\renewcommand{\t}{\theta}

\renewcommand{\l}{\lambda}

\renewcommand{\r}{\rho}
\newcommand{\s}{\sigma}
\renewcommand{\S}{\Sigma}

\newcommand{\mc}{\mathcal}

\newcommand{\mr}{\mathrm}
\newcommand{\mb}{\mathbb}

\newcommand{\bN}{\mb{N}}

\newcommand{\cB}{\mc{B}}

\newcommand{\cF}{\mc{F}}
\newcommand{\cG}{\mc{G}}

\newcommand{\cL}{\mc{L}}

\newcommand{\cN}{\mc{N}}

\newcommand{\cP}{\mc{P}}

\newcommand{\cR}{\mc{R}}
\newcommand{\cS}{\mc{S}}
\newcommand{\cT}{\mc{T}}

\newcommand{\cV}{\mc{V}}

\newcommand{\cX}{\mc{X}}

\newcommand{\vl}{{\vec{l}}}

\newcommand{\vp}{{\vec{p}}}
\newcommand{\vq}{{\vec{q}}}
\newcommand{\vr}{{\vec{r}}}
\newcommand{\vs}{{\vec{s}}}
\newcommand{\vt}{{\vec{t}}}
\newcommand{\vu}{{\vec{u}}}
\newcommand{\vv}{{\vec{v}}}

\newenvironment{rul}
  {$\begin{array}{rcl}}
  {\end{array}$}

\newenvironment{rew}[1][~~\a~~]
  {$\begin{array}{r@{#1}l}}
  {\end{array}$}
\newenvironment{rewc}[1][~~\a~~]
  {\begin{center}\begin{rew}[#1]}
  {\end{rew}\end{center}}

\newcounter{counter}

\newcounter{explnum}
{\theorembodyfont{\rmfamily} 
  \newtheorem{dfn}[counter]{Definition}
  \newtheorem{lem}[counter]{Lemma}
  \newtheorem{thm}[counter]{Theorem}
  \newtheorem{cor}[counter]{Corollary}

  \newtheorem{expl}[explnum]{Example}
}

\newcommand{\cqfd}{\hfill$\blacksquare$}

\newenvironment{prf}{{\bf Proof.}}{}

\newenvironment{lstgeneric}[2]
  {\begin{list}{#1}{\topsep=.5mm\itemsep=.5mm\parsep=0mm%
    \itemindent=-3ex\labelsep=1ex\labelwidth=0ex #2}}
  {\end{list}}

\newcommand{\SN}{{\cS\cN}}

\newcommand{\FrV}{{\cF\cV}}

\newcommand{\NF}{{\cN\cF}}
\newcommand{\trm}{{\cT\! rm}}

\newcommand{\Btree}{{\bs{B}}}

\newcommand{\lx}{\l x}

\newcommand{\lam}{\uplambda}

\newcommand{\col}{\colon\!}
\newcommand{\ceq}{\coloneqq}
\newcommand{\red}{\shortrightarrow}

\DeclareMathOperator{\match}{match}
\DeclareMathOperator{\size}{size}

\DeclareMathOperator{\leaf}{leaf}
\DeclareMathOperator{\Leaf}{Leaf}
\DeclareMathOperator{\node}{node}
\DeclareMathOperator{\Node}{Node}
\DeclareMathOperator{\app}{app}

\newcommand{\wild}{\_}

\newcommand{\subpat}{\ll}
\newcommand{\redmatch}{\subpat\shortdownarrow}

\newcommand{\len}[1]{|#1|}
\newcommand{\erase}{\len}
\newcommand{\efftrm}{\trm^{\erase{\centerdot}}}

\newcommand{\subst}[2]{\{#1\to#2\}}
\newcommand{\ext}[2]{^{#1}_{#2}}

\begin{document}

\title{Refinement Types as Higher-Order Dependency Pairs}
\author{Cody Roux}
\institute{INRIA-Nancy Grand Est}

\maketitle

\begin{abstract}
  Refinement types are a well-studied manner of performing in-depth
  analysis on functional programs. The dependency pair method is a
  very powerful method used to prove termination of rewrite systems;
  however its extension to higher-order rewrite systems is still the
  subject of active research. We observe that a variant of
  refinement types allows us to express a form of higher-order
  dependency pair method: from the rewrite system labeled with typing
  information, we
  build a \emph{type-level approximated dependency graph}, and
  describe a type level \emph{embedding-order}. We describe a
  syntactic termination criterion involving the graph and the order,
  and prove
  our main result: if the graph passes the criterion, then
  every well-typed term is strongly normalizing.
\end{abstract}

\section{Introduction}

Types are used to perform static analysis on programs. Various type
systems have been developed to infer information about termination,
run-time complexity, or the presence of uncaught exceptions.


We are interested in one such development, namely \emph{dependent
  types} \cite{McKinnaDepTy,DBAutomath}. Dependent types explicitly
allow ``object level'' terms to
appear in the types, and can express arbitrarily complex program
properties using the so called \emph{Curry-Howard
  isomorphism}. We are particularly interested here in
\emph{refinement types} \cite{XiScottDepTy,FreemanPfenningRefTy}.
For a given base type $B$ and a
property $P$ on programs, we may form a type $R$ which is a
\emph{refinement of $B$} and which is intuitively given the semantics: 
$$ R=\{t\col B\mid P(t)\}$$

Programing languages based on dependent type systems have the
reputation of being unwieldy, due to the perceived weight of proof
obligations in heavily specified types. The field of dependently typed
programing can be seen as a quest to find the compromise between
expressivity of types and ease of use for the programmer.

Dependency pairs are a highly successful technique for proving
termination of first-order rewrite systems
\cite{ArtsGieslDP}. However, without
modifications, it is difficult to apply the method to higher-order
rewrite systems. Indeed, the data-flow of such systems is
significantly different than that of first-order ones. Let us
examine the rewrite rule:

$$f\ (S\ x)\red (\l y.f\ y)\ x$$
The termination of well-typed terms under this rewrite system combined
with $\b$-reduction cannot be inferred by simply looking at
the left-hand side $f\ (S\ x)$ and the recursive call $f\ y$ in the right
hand side as it could be in first-order rewriting. Here we need
to infer that the variable $y$ can only be
instantiated by a subterm of $S\ x$. This can be done using dependent
types, using a framework called \emph{size-based termination} or
sometimes \emph{type-based termination}
\cite{hughes96popl,abel04ita,barthe04mscs,blanqui04rta,blanqui06lpar-cbt}.

The dependency pair method rests on the examination of the aptly-named
\emph{dependency pairs}, which correspond to left-hand sides of rules
and function calls with their arguments in the right-hand side of the
rules. For instance with a rule
$$f(c(x,y),z)\red g(f(x,y)) $$
We would have two dependency pairs, the pair $f(c(x,y),z)\red f(x,y)$
and the pair $f(c(x,y),z)\red g(f(x,y))$.

We can then define a \emph{chain} to be a pair $(\t,\phi)$ of
substitutions, and a couple\linebreak $(t_1\red u_1,t_2\red
  u_2)$ of
dependency pairs such that $u_1\t\red^* t_2\phi$. We may connect chains in
an intuitive manner, and the fundamental
theorem of dependency pairs may be stated: \emph{a (first-order)
  rewrite system is terminating if and only if there are no infinite
  chains}. See also the original article \cite{ArtsGieslDP} for
details.

To prove that no infinite chains exist, one wants to work with the
\emph{dependency graph}: the
graph built using the dependency pairs as nodes and with a
vertex between $N_1=t_1\red u_1$ and $N_2=t_2\red u_2$ if
there exist 
$\t$ and $\phi$ such that $(\t,\phi),(N_1,N_2)$ form a chain. It is
then shown that if the system is finite, then it is sufficient to
consider only the cycles in this
graph and prove that they may not lead to infinite chains
\cite{giesl02jsc}.
It is known that in general computing the dependency graph is
undecidable (this is the \emph{unification modulo rewriting}
problem, see \eg Jouannaud \etal. \cite{JouannaudUnif}), so in practice
we compute an approximation (or estimation) of the graph that
is \emph{conservative}: all edges in the dependency graph are sure to
appear in the approximated graph. One common (see for instance Giesl
\cite{giesl06jar}) and reasonable approximation is to perform ordinary
unification on non-defined symbols (that is, symbols that are not at
the head of a left-hand side), while replacing each subterm headed by
a defined symbol by a fresh variable, ensuring that it may unify with
any other term.

In this article, we show that the dependency pair technique with the
approximated dependency graph can be modeled using a form of
refinement types containing \emph{patterns} which denote sets of
possible values to which a term reduces. These type-patterns must be
explicitly abstracted and applied, a choice that allows us to have
very simple type inference. This allows
us to build a notion of type-based dependency pair for higher-order
rewrite rules, as well as an approximated dependency graph which
corresponds to the estimation described
above. We describe an order on the type annotations, that essentially
capture the subterm ordering, and use this order to express a
\emph{decrease condition} along cycles in the approximated dependency
graph. We then state the correctness of the
criterion: if in every \emph{strongly connected component} of the
graph and every cycle in the component, the decrease condition holds,
then every well-typed term is strongly normalizing under the rewrite
rules and $\b$-reduction.
The actual operational semantics are defined not on the terms
themselves, but on \emph{erased terms} in which we remove the explicit
type information. We then
conclude with a comparison with other approaches to higher-order
dependency pairs and possible extensions of our criterion.

\section{Syntax and Typing Rules}

The language we consider is simply a variant of the $\l$-calculus with
constants. For simplicity we only consider the datatype of binary
(unlabeled) trees. The development may be generalized without
difficulty to other first-order datatypes, \ie types whose
constructors do not have higher-order recursive arguments. We define
the syntax of \emph{patterns}

\[p,q\in\cP\ceq \al\mid \leaf\mid \node(p,q)\mid\wild\mid\bot  \]
With $\al\in\cV$ a set of \emph{pattern variables}, and $\_$ is called
\emph{wildcard}. Patterns appear in types to describe possible
reducts of terms. We define the set of types:

\[T,U\in\cT\ceq \Btree(p)\mid T\a U\mid \all\al.T\]

An \emph{atomic type} is a type of the form $\Btree(p)$.
The set of terms of our language is defined by:

\[t,u\in\trm\ceq x\mid f\mid t\ u\mid t\ p\mid \lx\col T.t\mid \lam
\al.t\mid \Node\mid \Leaf \]
With $x\in\cX$ a set of term variables, $f\in\S$ is a set of
\emph{function symbols} and $\al\in\cV$. Defined symbols are in lower case.
Notice that application and abstraction of patterns is explicit. A
\emph{constructor} is either $\Node$ or $\Leaf$.
A \emph{context} is a list of judgements $x\col T$ with $x\in\cX$ and
$T\in\cT$, with each variable appearing only once.

Intuitively, $\Btree(p)$ denotes the set of terms that
reduce to some term that \emph{matches} the pattern $p$. For instance,
any binary tree $t$ is in the semantics of $\Btree(\wild)$, only
binary trees that reduce to $\Node\ t_1\ t_2$ for some binary trees
$t_1$ and $t_2$ are in $\Btree(\node(\wild,\wild))$, and only terms
that \emph{never} reduce to a constructor are in $\Btree(\bot)$. Our
operational semantics are defined by rewriting, which has the
following consequences, which may be surprising to a programming
language theorist:

\begin{itemize}
\item It may be the case that a term $t$ has several distinct normal
  forms. Indeed we do not require our system to be orthogonal, or even
  confluent (we do require it to be finitely branching
  though). Therefore a term is in the semantics of
  $\Btree(\node(\wild,\wild))$ if \emph{all} its reducts reduce to a
  term of the form $\Node\ t\ u$.
\item It is possible for a term to be \emph{stuck} in the empty
  context, that is in normal form and not headed by a constructor or
  an abstraction. Therefore $\Btree(\bot)$ is not necessarily empty
  even in the empty context.
\end{itemize}

We write
$\cF\cV(t)$ (resp. $\cF\cV(T)$, $\cF\cV(\G)$) for the set of free
variables in a term $t$ (resp. a type $T$, a context $\G$). If a term
(resp. pattern) does not contain any free variables, we say that it is
\emph{closed}. We write
$\all\vec{\al}.T$ for $\all\al_1.\all\al_2\ldots\all\al_n.T$, and
arrows and
application are associative to the left and right respectively, as
usual. A pattern variable $\al$ appears in $\Btree(p)$ if it appears
in $p$. It appears \emph{positively} in a type $T$ if:

\begin{itemize}
\item $T=\Btree(p)$ and $\al$ appears in $p$
\item $T=T_1\a T_2$ and $\al$ appears positively in $T_2$ or
  negatively in $T_1$ (or both).
\end{itemize}

With $\al$ appearing \emph{negatively} in $T$ if $T=T_1\a T_2$ and
$\al$ appears negatively in $T_2$ or positively in $T_1$ (or both).

We consider a \emph{type assignment} $\tau\col\S\a \cT$, such that
for each $f\in\S$, there is a number $k$ such that
$\tau_f=\all\al_1,\ldots,\al_n.A_1\a\ldots\a A_k\a T_f$ with
\begin{itemize}
\item $n\geq k$
\item $A_i=\Btree(\al_i)$
\item $\all 1\leq i\leq k, \al_i$ appears \emph{positively} in $T_f$.
\end{itemize}
In this case $k$ is called the number of \emph{recursive arguments}.

The positivity condition is quite similar to the one used in the usual
formulation of type-based termination, see for instance
Abel \cite{DBLP:conf/csl/Abel06} for an in depth analysis. The typing rules are
also similar to the ones for type-based termination. The typing rules
of our system are given by the typing rules in figure \ref{fig:type_rules}.

\begin{figure}
  
  \begin{prooftree}
    \AxiomC{}
    \RightLabel{\bf ax}
    \UnaryInfC{$\G,x\col T,\D\th x\col T$}
  \end{prooftree}
  
  \begin{prooftree}
    \AxiomC{$\G,x\col T\th t\col U$}
    \RightLabel{\bf t-lam}
    \UnaryInfC{$\G\th\lx\col T.t\col T\a U$}
  \end{prooftree}

  \begin{prooftree}
    \AxiomC{$\G\th t\col T$}
    \RightLabel{\bf p-lam}
    \LeftLabel{$\al\notin\cF\cV(\G)$}
    \UnaryInfC{$\G\th\lam\al.t\col\all\al.T$}
  \end{prooftree}

  \begin{prooftree}
    \AxiomC{}
    \RightLabel{\bf leaf-intro}
    \UnaryInfC{$\G\th \Leaf\col\bs{B}(\leaf)$}
  \end{prooftree}
  
  \begin{prooftree}
    \AxiomC{}
    \RightLabel{\bf node-intro}
    \UnaryInfC{$\G\th\Node\col\all\al\b.\bs{B}(\al)\a\bs{B}(\b)\a\bs{B}(\node(\al,\b))$}
  \end{prooftree}
  
  \begin{prooftree}
  \AxiomC{$\G\th t\col T\a U$}
  \AxiomC{$\G\th u\col T$}
  \RightLabel{\bf t-app}
  \BinaryInfC{$\G\th t\ u\col U$}
  \end{prooftree}

  \begin{prooftree}
    \AxiomC{$\G\th t\col\all \al.T$}
    \RightLabel{\bf p-app}
    \UnaryInfC{$\G\th t\ p\col T\subst{\al}{p}$}
  \end{prooftree}
  
  \begin{prooftree}
    \AxiomC{}
    \RightLabel{\bf symb}
    \UnaryInfC{$\G\th f\col\tau_f$}
  \end{prooftree}
  
  \caption{Typing Rules}
  \label{fig:type_rules}

\end{figure}

To these rules we add the subtyping rule:

\begin{prooftree}
  \AxiomC{$G\th t\col T$}
  \AxiomC{$T\leq U$}
  \RightLabel{\bf sub}
  \BinaryInfC{$\G\th t\col U$}
\end{prooftree}

Where the subtyping relation is defined by an order on patterns:

\begin{itemize}
\item $p\subpat \wild$
\item $\al\subpat \al$
\item $\node\subpat\node$
\item $p_1\subpat q_1\et p_2\subpat q_2\A\node(p_1,p_2)\subpat\node(q_1,q_2)$
\item $\bot\subpat p$
\end{itemize}
For all patterns $p,p_1,p_2,q_1,q_2$. This order is carried to types by:

\begin{itemize}
\item $p\subpat q\A \bs{B}(p)\leq\bs{B}(q)$
\item $T_2\leq T_1\et U_1\leq U_2\A T_1\a U_1\leq T_2\a U_2$
\item $T\leq U \A\all \al.T\leq\all \al.U$
\end{itemize}

This type system is quite similar to the refinement types described
for mini-ML by Freeman \etal. \cite{FreemanPfenningRefTy}, and is
not very distant from \emph{generalized algebraic
  datatypes} as are implemented in certain Haskell compilers
\cite{Jones06simpleunification-based}, though subtyping is not present
in that framework.

It may seem surprising that we choose to explicitly represent pattern
abstraction and application in our system. This choice is justified by
the simplicity of type inference with explicit parameters. In the
author's opinion, implicit arguments should be handled by the
following schema: at the user level a language without
implicit parameters; these parameters are inferred by the compiler,
which type-checks a language with all parameters
present. Then at run-time they are once again erased. This
is exactly analogous to a Hindley-Milner type language in which System
F is used as an intermediate language
\cite{Milner78atheory,JonesMeijerHenk}.
It is also our belief that explicit parameters will allow this
criterion to be more easily integrated into languages with
pre-existing dependent types, \eg {Adga} \cite{norell:thesis},
{Epigram} \cite{McKinnaDepTy} or {Coq} \cite{coq}.

A \emph{constructor term} $l\in\cL$ is a term built following the
rules:
\[l_1,l_2\in\cL\ceq x\mid \Leaf\mid \Node\ l_1\ l_2 \]
with $x\in\cX$.

A rewrite rule is a pair of terms $(l,r)$ which we write $l\red r$,
such that $l$ is of the form $f\ p_1\ldots p_n\ l_1\ldots l_k$ with
$f\in\S$, $p_i\in\cP$ and $l_i\in\cL$, such that $k$ is the number of
recursive arguments of $f$. We suppose that the free
variables of $r$ appear in $l$.

We suppose in addition that every function symbol $g\in r$ is
\emph{fully applied} to its pattern arguments, that is if
$\tau_g=\all\al_1\ldots \al_l.T$ then for each occurrence of $g$ in $r$
there are patterns $p_1,\ldots,p_l\in\cP$ such that $g\ p_1\ldots
p_l$ appears at that position.

In the following we consider a \emph{finite} set $\cR$ of rewrite
rules. The set $\cR$ is \emph{well-typed} if for each rule $l\red
r\in\cR$, there is a context $\G$ and a type $T$ such that
\[\G\th_{min}l\col T \]
and
\[\G\th r\col T \]
with $\th_{\min}$ defined in
figure \ref{fig:min_type_rules}.

\begin{figure}
  
  \begin{prooftree}
    \AxiomC{}
    \RightLabel{$\al\notin\G,\G'$}
    \UnaryInfC{$\G,x\col\bs{B}(\al),\G'\th_{\min} x\col\bs{B}(\al)$}
  \end{prooftree}

  \begin{prooftree}
    \AxiomC{}
    \RightLabel{}
    \UnaryInfC{$\G\th_{\min}\Leaf\col\bs{B}(\leaf)$}
  \end{prooftree}
  
  \begin{prooftree}
    \AxiomC{$\G\th_{\min} l_1\col\bs{B}(p_1)$}
    \AxiomC{$\G\th_{\min} l_2\col\bs{B}(p_2)$}
    \RightLabel{}
    \BinaryInfC{$\G\th_{\min}\Node\ p_1\ p_2\ l_1\ l_2\col
      \bs{B}(\node(p_1,p_2))$}
  \end{prooftree}

  \begin{prooftree}
    \AxiomC{$\G\th_{\min} l_1\col\bs{B}(p_1)$}
    \AxiomC{$\ldots$}
    \AxiomC{$\G\th_{\min}  l_k\col\bs{B}(p_k)$}
    \RightLabel{$\vec{\al}\notin\G$}
    \TrinaryInfC{$\G\th_{\min} f\ p_1\ldots
      p_k\ \b_{k+1}\ldots\b_l\ l_1\ldots l_k\col T_f\phi$}
  \end{prooftree}
  
  With $\tau_f=\all\al_1\ldots\al_l. A_1\a\ldots\a A_k\a T_f$ and
  $\phi(\al_i)=p_i$ if $1\leq i\leq k$ and $\phi(\al_j)=\b_j$ for
  $k<i\leq l$.
  \caption{Minimal Typing Rules}
  \label{fig:min_type_rules}
\end{figure}

Notice that if $\G\th_{\min} l_i\col T$ then $T$ is
\emph{unique}. Minimal typing is present in other work on
size-based termination \cite{DBLP:conf/csl/BlanquiR09}, in which it is
called the \emph{pattern condition}. The purpose of minimal typing is to
constrain the possible types of constructor terms in left hand
sides.


We can then define the higher-order analogue of dependency pairs,
which use type information instead of term information.

\begin{dfn}
  Let $\r= f\ \vp\ \vl\red r$ be a rule in $\cR$, with $\G$ such
  that $\G\th_{min} f\ \vp\ \vl\col T$, and $\G\th r\col T$. The set
  of \emph{type dependency pairs} $DP_\cT(\r)$ is the set
  \[\{f^\sharp(p_1,\ldots,p_k)\red g^\sharp(q_1,\ldots,q_l)\mid
  \all i, \G\th_{\min} l_i\col \Btree(p_i)
  \et g\ q_1\ldots q_l\mbox{ appears in }r\}\]
  The set $DP_\cT(\cR)$ is defined as the union of all $DP_\cT(\r)$,
  for $\r\in\cR$, where we suppose that all variables are disjoint
  between dependency pairs.
\end{dfn}

The set of higher-order dependency pairs defined above should already
be seen as an abstraction of the traditional dependency pair notion
(for example those defined in \cite{ArtsGieslDP}). Indeed, due to
subtyping, there may be some information lost in the types, if for
instance the wildcard pattern is used. As an example, if $f, g$ and $h$ all
have type $\all\al.\Btree(\al)\a\Btree(\wild)$, consider the rule
\[f\ \al\ x\red g\ \wild\ (h\ \al\ x) \]
The dependency pair we obtain is
\[f^\sharp(x)\red g^\sharp(\wild) \]
The information that $g$ is called on the argument $h\ x$ is
lost.

This approach can therefore be seen as a type based manner to
study an approximation of the dependency graph. Note that in the
case where $h$ is given a more precise type, like
$\Btree(\al)\a\Btree(\leaf)$, which is the case if every normal
form of $h\ t$ is either neutral or $\Leaf$, we have a more precise
approximation.

Note that, in addition, a dependency pair is not formally a
(higher-order) rewrite rule, though it may be seen as a first-order
one.

\begin{dfn}
  Let $p$ and $q$ be patterns. We say that $p$ and $q$ are
  \emph{pattern-unifyable}, and write $p\bowtie q$, if $p'$ and $q'$
  are unifyable, where $p'$ and $q'$ are the patterns $p$ and $q$ in
  which each occurrence of $\wild$ and each occurrence of a variable is
  replaced by some \emph{fresh} variable.

  The \emph{standard typed dependency graph} $\cG_\cR$ is defined as
  the graph with
  
  \begin{itemize}
  \item As set of nodes the set $DP_\cT(\cR)$.
  \item An edge between the dependency pairs $t\red
    g^\sharp(p_1,\ldots,p_k)$ and $h^\sharp(q_1,\ldots,q_l)\red u$ if
    $g=h$, $k=l$ and for every $ 1\leq i\leq k, p_i\bowtie q_i$.
  \end{itemize}
\end{dfn}

This definition gives us an adequate higher-order notion of standard
approximated dependency graph. We will now show that it is possible to
give an order on the terms in the dependency pairs, which is
similar to a simplification order and which will allow us to show
termination of well-typed terms under the rules, if the graph
satisfies an intuitive decrease criterion.

\begin{dfn}
  We define the \emph{embeddeding preorder} on $\cP$ written $p\tgt q$ by
  the following rules
  
  \begin{itemize}
  \item $p_i\tge q \A \node(p_1,p_2)\tgt q$ for $i=1,2$
  \item $p_1\tgt q_1\et p_2\tge
    q_2\A \node(p_1,p_2)\tgt\node(q_1,q_2)$
  \item $p_1\tge q_1\et p_2\tgt
    q_2\A \node(p_1,p_2)\tgt\node(q_1,q_2)$
  \end{itemize}
  With $\tge$ as the reflexive closure of $\tgt$ and
  with the further condition that if $p\tgt q$, then $p$ and $q$ may
  not contain any occurrence of $\wild$.

\end{dfn}

Non termination can intuitively be traced to cycles in the dependency
graph. We wish to consider termination on terms with erased pattern
arguments and type annotations.

\section{Operational Semantics and the Main Theorem}

Rewriting needs to be performed over terms with erased pattern
annotations. The problem with the na\"\i ve definition of rewriting
arises when trying to match on
patterns. Take the rule \[f\ \node(\al,\b)\ (\Node\ x\
y)\red \Leaf\] In the presence of this rule, we wish to have, for
instance, the
reduction \[f\ \wild\ (\Node\ (g\ x)\ (h\ x))\red \Leaf\] However,
there is no substitution $\t$ such that
$\node(\al,\b)\t=\wild$. There are two ways to deal with this. Either
we take subtyping into account when performing matching, or we do away
with the pattern arguments when performing reduction. We adopt the
second solution, as it is used in practice when dealing with languages
with dependent type annotations (see for example McKinna
\cite{McKinnaDepTy}). Symmetrically, we erase pattern abstractions as
well.

\begin{dfn}
  We define the set of \emph{erased terms} $\efftrm$ as:
  $$t,u\in\efftrm\coloneqq
  x\mid f\mid\lx.t\mid t\ u\mid\Leaf\mid\Node$$
  Where $x\in\cX$ and $f\in\cF$.
  
  Given a term $t\in\trm$, we define the \emph{erasure}
  $\erase{t}\in\efftrm$ of $t$ as:
  \[
  \begin{array}{l@{~=~}l}
  \erase{x} & x\\
  \erase{f} & f\\
  \erase{\lx\col T.t} & \lx.\erase{t}\\
  \erase{\lam\al.t} & \erase{t}\\
  \erase{t\ u} & \erase{t}\ \erase{u}\\
  \erase{t\ p} & \erase{t}\\
  \erase{\Leaf} & \Leaf\\
  \erase{\Node} & \Node
  \end{array}
  \]
\end{dfn}

An erased term can intuitively be thought of as the compiled form of a
well typed term.

\begin{dfn}
  An erased term $t$ \emph{head rewrites} to a term $u$ if there is
  some rule 
  $l\red r\in\cR$ and some substitution $\s$ from $\cX$ to terms in
  $\efftrm$ such that
  \[\erase{l}\s=t\et \erase{r}\s=u\]
  We define $\b$-reduction $\red_{\b}$ as 
  \[\lx.t\ u\red_\b t\subst{x}{u} \]
  And we define the \emph{reduction} $\red$ as the closure of
  head-rewriting and $\b$-reduction by term contexts.
  We then define $\red^*$ and $\red^+$ as the
  symmetric transitive and transitive closure of $\red$,
  respectively.
\end{dfn}

We can now express our termination criterion. We need to consider the
\emph{strongly connected components}, or SCCs of the typed dependency
graph. A strongly connected component of a graph $\cG$ is a full
subgraph such that each node is reachable from all the others.

\begin{thm}\label{ho_dp:thm:sn}
  Let $\cG$ be the typed dependency graph for $\cR$ and let
  $\cG_1,\ldots,\cG_n$ be the SCCs of $\cG$. Suppose that
  for each $\cG_i$, there is a \emph{recursive
    index} $\iota^i\col \S\a \bN$ which to
  $f\in\S$ associates
  an integer $1\leq\iota^i_f\leq k$ (with $k$ the number of recursive
  arguments of $f$).

  Suppose that for each $1\leq i\leq n$ and each rule
  $f^\sharp(p_1,\ldots,p_n)\red g^\sharp(q_1,\ldots,q_m)$ in $\cG_i$,
  we have $p_{\iota^i_f}\tge q_{\iota^i_g}$. Finally suppose that for
  each cycle in $\cG_i$, there is some rule
  $f^\sharp(p_1,\ldots,p_n)\red g^\sharp(q_1,\ldots,q_m)$ such that
  \[p_{\iota^i_f}\tgt q_{\iota^i_g} \]
  then for every $\G,t,T$ such that $\G\th t\col T$,
  \[\erase{t}\in\SN_\cR \]
\end{thm}

The proof of this theorem can be found in the appendix.
Let us give two examples of the application of this technique.

\begin{expl}\label{expl:app}
Take the
rewrite system given by the signature:
$\{\app\col\all\al\b.(\Btree(\al)\a\Btree(\b))\a\Btree(\al)\a\Btree(\b),
f\col\Btree(\leaf), g\col\all\al.\Btree(\al)\a\Btree(\leaf)\}$,
. We give the rewrite
rules:
\[\app\ \a \lam\al\b.\lx\col\Btree(\al)\a\Btree(\b).\l y\col\Btree(\al). x\ y \]
\[f \a
\app\ \node(leaf,leaf)\ \leaf
\ (g\ \node(\leaf,\leaf))\ (\Node\leaf\leaf\ \Leaf\ \Leaf)\]
\[g\ \node(\al,\b)\ (\Node\al\ \b\ x\ y) \a \Leaf \]
\[g\ \leaf\ \Leaf \a f \]

or, in more readable form with pattern arguments and type annotations
omitted:

\[
\begin{array}{r@{~\a~}l}
  \app & \lx.\l y.x\ y\\
  f & \app\ g\ (\Node\ \Leaf\ \Leaf)\\
  g\ (\Node\ x\ y) & \Leaf\\
  g\ \Leaf & f\\
\end{array}
\]
It is possible to verify that the criterion can be applied and that
in consequence, according to theorem \ref{ho_dp:thm:sn}, all
well typed terms are strongly normalizing under $\cR\cup\b$.
\end{expl}

\begin{figure}
\begin{tikzpicture}
  \node (n1) at (0,0) {$g^\sharp(\leaf)\red f^\sharp$};
  \node (n2) at (6,2) {$f^\sharp\red \app^\sharp$};
  \node (n3) at (6,-2) {$f^\sharp\red g^\sharp(\node(\leaf,\leaf))$};
  \draw[->] (n1) to [out=30,in=180]  (n2);
  \draw[->] (n1) to [out=-30,in=180] (n3);
\end{tikzpicture}
\caption{Dependency graph of example \ref{expl:app}}
\label{fig:dep_grph}
\end{figure}
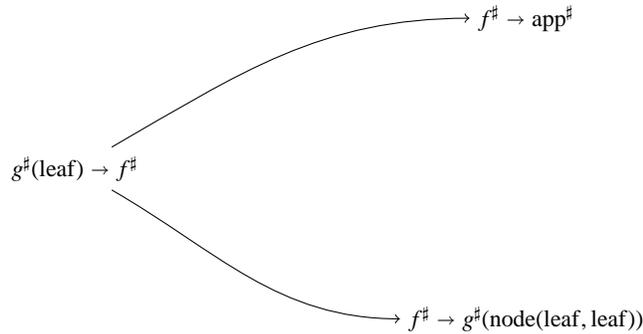

Indeed, we may easily check that each of these rules is minimally
typed in
some context. Furthermore, we can check that the dependency graph in
figure \ref{fig:dep_grph} has
no cycles.

One may object that if we inline the definition of $\app$ and perform
$\b$-reduction on the right-hand sides of rules we obtain a rewrite
system that can be treated with more conventional methods, such as
those performed by the \texttt{AProVe} tool \cite{Giesl05provingand}
(on terms without abstraction, and
without $\b$-reduction). However this operation can be
very costly if performed automatically and is, in its most na\"\i ve
form, ineffective for even slightly more complex higher-order programs
such as $map$, which performs pattern matching and for which we need
to instantiate. By resorting to typing, we allow termination to be
proven using only ``local'' considerations, as the information
encoding the semantics of $\app$ is contained in its type.

However it becomes
necessary, if one desires a fully automated termination check on an
unannotated system, to somehow infer the type of defined constants,
and possibly perform an analysis quite similar in effect to the one
proposed above. We believe that to this end one may apply known type
inference technology, such as the one described in \cite{chin01hosc},
to compute these annotated types. In conclusion, what used to be a
termination problem becomes a type inference problem, and may benefit
from the knowledge and techniques of this new community, as well as
facilitate integration of these techniques into type-theoretic based
proof assistants like {Coq} \cite{coq}.

Let us examine a second, slightly more complex example, in which there
is ``real'' recursion.

\begin{figure}\label{fig:dp_expl2}

\begin{tikzpicture}[node distance=1.4]
  \node (i1) at (0,0) {$i^\sharp(\node(\al,\b))\red i^\sharp(\al)$};
  \node (i2) [below=of i1] {$i^\sharp(\node(\al,\b))\red 
    i^\sharp(\b)$};
  \node (f1) [right = of i1] {$f^\sharp(\node(\al,\b))\red g^\sharp(\node(\al,\b))$};
  \node (f2) [below = of f1] {$f^\sharp(\node(\al,\b))\red i^\sharp(\node(\al,\b))$};
  \node (g1) [right = of f1] {$g^\sharp(\node(\al,\b))\red f^\sharp(\al)$};
  \node (g2) [below = of g1] {$g^\sharp(\node(\al,\b))\red i^\sharp(\node(\al,\b))$};
  \node (g3) [below = of f2] {$g^\sharp(\leaf)\red f^\sharp(\bot)$};
  \node (g4) [right = of g3] {$g^\sharp(leaf)\red h^\sharp(\leaf)$};
  \node (h1) [below = of i2] {$h^\sharp(\node(\al,\b))\red h^\sharp(\al)$};

  \path[->] (i1) edge [bend right] (i2);
  \path[->] (i2) edge [bend right] (i1);
  \path[->] (i1) edge [loop above] (i1);
  \path[->] (i2) edge [loop below] (i2);
  \path[->] (f1) edge [bend right] (g1);
  \draw[->] (f1) to [out=-60,in=178] (g2);
  \path[->] (g1) edge [bend right] (f1);
  \draw[->] (g1) to [out=-120,in=10] (f2);
  \path[->] (g2) edge [bend right] (f2);
  \draw[->] (f2) to [out=170,in=-45] (i1);
  \path[->] (f2) edge              (i2);
  \path[->] (g2) edge              (i1);
  \path[->] (g2) edge [bend left]  (i2);
  \path[->] (h1) edge [loop above] (h1);
\end{tikzpicture}
\caption{The dependency graph for example \ref{expl:decrease}}
\end{figure}
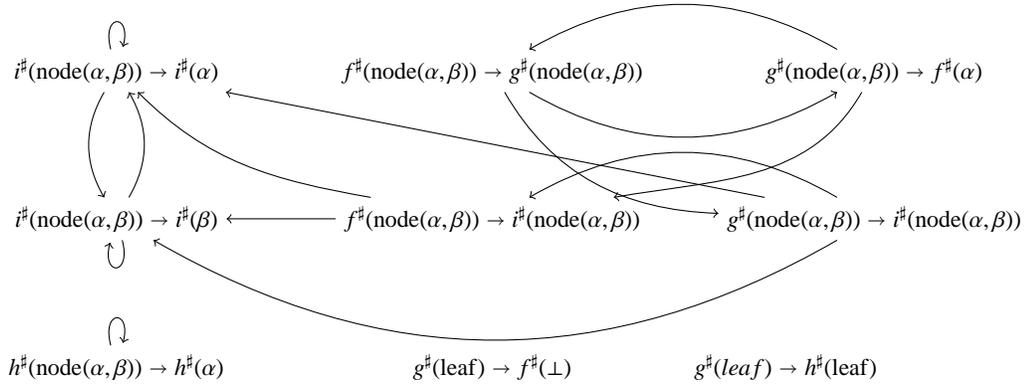

\begin{expl}\label{expl:decrease}
  Let $\cR$ be the rewrite system defined by
  \[
  \begin{array}{r@{~\a~}l}
    f\ (\Node\ x\ y) & g\ (i\ (\Node\ x\ y)\\
    g\ (\Node\ x\ y) & f\ (i\ x)\\
    g\ \Leaf & f\ (h\ \Leaf)\\
    i\ (\Node\ x\ y) & \Node\ (i\ x)\ (i\ y)\\
    i\ \Leaf & \Leaf\\
    h\ (\Node\ x\ y) & h\ x
  \end{array}
  \]
  Again with the type arguments omited, and with types
  $f,g\col\all\al.\Btree(\al)\a \Btree(\wild)$, $h\col
  \all\al.\Btree(\al)\a \Btree(\bot)$ and
  $i\col\all\al.\Btree(\al)\a\Btree(\al)$. Every equation can by typed
  in the context $\G=x\col \Btree(\al),y\col \Btree(\b)$, The system
  with full type annotations is given in the appendix.
 
  The dependency graph is given in figure \ref{fig:dp_expl2}, and has
  as SCCs the full subgraphs of $\cG_\cR$ with nodes
  $\{i^\sharp(\node(\al,\b))\red i^\sharp(\al),
  i^\sharp(\node(\al,\b))\red i^\sharp(\b)\}$,
  $\{f^\sharp(\node(\al,\b))\red g^\sharp(\node(\al,\b),
  g^\sharp(\node(\al,\b)\red f^\sharp(\al)\}$
  and $\{h^\sharp(\node(\al,\b))\red h^\sharp(\al)\}$
  respectively. 

  Taking $\iota_s=1$ for every SCC and every symbol
  $s\in\S$, it is easy to show
  that every SCC respects the decrease criterion on cycles. For
  example, in the cycle 
  \[f^\sharp(\node(\al,\b))\red
  g^\sharp(\node(\al,\b)) \leftrightarrows g^\sharp(\node(\al,\b))\red
  f^\sharp(\al)\]
  we have $\node(\al,\b)\tge\node(\al,\b)$ and
  $\node(\al,\b)\tgt\al$, so the cycle is weakly decreasing with at
  least one strict decrease.

  We may then again apply the correctness theorem to conclude that the
  erasure of all well-typed terms are strongly normalizing with
  respect to $\cR\cup\b$.

\end{expl}

Note that the minimality condition is important: otherwise one could
take $f\col\all \al\b.\Btree(\al)\a\Btree(\b)\a\Btree(\_)$ with the
rule

\[ f \node(\leaf,\leaf) \leaf\ x\ y \red f \leaf \leaf\ y\ y\]
This rule can be typed in the context $x\col
\Btree(\node(\leaf,\leaf)),y\col \Btree(\leaf)$, but not minimally
typed, and passes the termination criterion: the dependency graph is
without cycles, as $\node(\leaf,\leaf)$ does not unify with
$\leaf$. However, this system leads to the non terminating reduction
$f \Leaf \Leaf \red f \Leaf \Leaf$.

\section{Comparison, future work}

Several extensions of dependency pairs to different forms of
higher-order rewriting have been proposed
\cite{BlanquiHODP,blanqui06wst-hodp,Giesl05provingand,Sakai_KusakariHODP,DBLP:conf/rta/AotoY05}.
However, these frameworks do not handle the presence of bound
variables, for which the usual approach is to defunctionalize (also
called \emph{lambda-lifting}) 
\cite{DBLP:conf/ppdp/DanvyN01,Johnsson85lambdalifting}.

In particular, all the techniques cited above, when
applied to example \ref{expl:app}, where we replace the rule
$\app\a\lx.\l y. x\ y$ with the rule
$\app\ x\ y\a x\ y$ (which does not involve bound variables), generate
a dependency graph with cycles. For example, in Sakai \& Kusakari
\cite{Sakai_KusakariHODP}, using the SN framework the dependency graph
is:

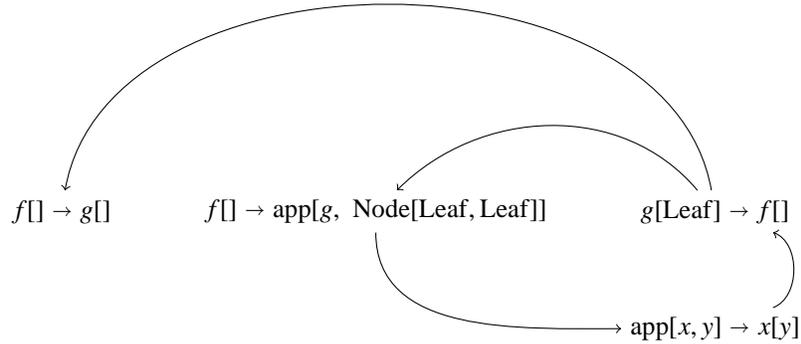
\begin{tikzpicture}
  \node (n1) at (0,0) {$f[]\red g[]$};
  \node (n2) [right=of n1] {$f[]\red\app[g,\ \Node[\Leaf, \Leaf]]$};
  \node (n3) [right=of n2] {$g[\Leaf]\red f[]$};
  \node (n4) [below=of n3] {$\app[x, y]\red x[y]$};
  \draw[->] (n2) to [out=-90,in=180]  (n4);
  \draw[->] (n4) to [out=20,in=-20] (n3);
  \draw[->] (n3) to [out=130,in=45] (n2);
  \draw[->] (n3) to [out=100,in=80] (n1);
\end{tikzpicture}

It is of course possible to prove that there are no infinite chains
for this problem (the criterion is complete), but we have not much
progressed from the initial
formulation!

Using the SC-framework from the same paper, which is based on
computability (as is our
framework), we obtain the following graph:

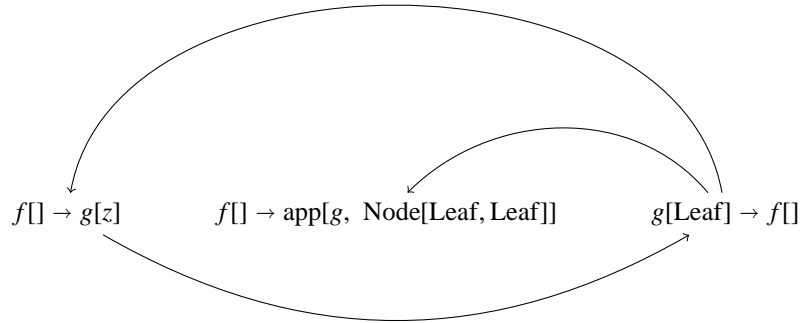
\begin{tikzpicture}
  \node (n1) at (0,0) {$f[]\red g[z]$};
  \node (n2) [right=of n1] {$f[]\red \app[g,\ \Node[\Leaf, \Leaf]]$};
  \node (n3) [right=of n2] {$g[\Leaf]\red f[]$};
  \draw[->] (n1) to [out=-30,in=-150]  (n3);
  \draw[->] (n3) to [out=130,in=45] (n2);
  \draw[->] (n3) to [out=100,in=80] (n1);
\end{tikzpicture}

However it is not possible to prove that there are no infinite chains
for this problem, as there is one! Therefore the criterion presented
in this paper allows a finer analysis of the possible calls.

The termination checking software \texttt{AProVE}
\cite{Giesl05provingand} succeeds in proving termination of example
\ref{expl:app}, by using an analysis involving instance computation
and symbolic reduction. As noted previously, it
seems that such an analysis may be used to infer the type
annotations required in our framework. At the moment it is unclear how
the typing approach compares to these techniques. More investigation
is clearly needed in this direction.

\texttt{AProVE} can also easily prove termination of the second
rewrite system (example \ref{expl:decrease}). However semantic
information needs to be inferred (for example a polynomial
interpretation needs to be given) when trying to well-order the cycle
\[f\ (\Node\ x\ y) \red g\ (i\ (\Node\ x\ y)\leftrightarrows
g\ (\Node\ x\ y)\red f\ (i\ x) \]
This information is already supplied by our type system (through the fact
that $i$ is of type $\forall \al.\Btree(\al)\a \Btree(\al)$), and
therefore it suffices to consider only syntactic information on the
approximated dependency graph. The \emph{subterm criterion} by Aoto
and Yamada \cite{DBLP:conf/rta/AotoY05} is insufficient to treat this
example.

The framework described here is only the first step towards a
satisfactory higher-order dependency pair framework using refinement
types. We intuitively consider a ``type level'' first-order rewrite
system, use standard techniques to show that that system is
terminating, and show that this implies termination of the object
level system. More work is required to obtain a satisfactory
``dependency pairs by typing'' framework.

Our work seems quite
orthogonal to the \emph{size-change principle}
\cite{Lee01thesize-change}, which suggests we could apply this
principle to treat cycles in the typed dependency graph, as a more
powerful criterion than simple decrease on one indexed argument.

It is clear that the definitions and proofs in the current work extend
to other first-order inductive types like lists, Peano natural
numbers, etc. We conjecture that this 
framework can be extended to more general positive inductive
types, like the type of Brower ordinals
\cite{blanqui02tcs}. These kinds of inductive
types seem to be difficult to treat with other (non type-based) methods.

For now types have to be explicitly given by the user, and it would
be interesting to investigate inference of annotations. Notice that
trivial
annotations (return type always $\Btree(\_)$) can very easily be infered
automatically. Some work on automatic inference of type-level
annotations has been carried out by Chin \etal. \cite{chin01hosc} which
may provide inspiration. On the other hand, we believe that the
inference of the explicit type information in the terms is quite
feasible with current state-of-the-art methods, for example those
used for inferring the type of functional programs using GADTs
\cite{Jones06simpleunification-based}.

We believe that refinement types are simply an alternative way of
presenting the dependency pair method for higher-order rewrite
systems. It is the occasion to draw a parallel between the types
community and the rewriting community, by emphasizing that techniques
used for the inference of dependent type annotations (for example work
on \emph{liquid types} \cite{DBLP:conf/pldi/RondonKJ08}), may in fact
be used to infer information necessary for proving termination and
(we believe) vice-versa. It may also be interesting in the case of a
programming language for the user to supply the types as
documentation, in what some call ``type directed programing''.

We only consider matching on non-defined symbols, though an extension
to a framework with matching on defined symbols seems feasible if we
add some conversion rule to our type system.

\subsection*{Acknowledgements}
We thank Frederic Blanqui for the discussions that led to the birth
of this work and for very insightful comments concerning a draft of
this paper, as well as anonymous referees for numerous corrections on a
previous version of this paper.

\bibliographystyle{alpha}
\bibliography{biblio}

\appendix

\section{The full system of example \ref{expl:decrease}}

Every rule is typed in the context $x\col \Btree(\al),y\col
\Btree(\b)$, and we remind that the types of defined functions are:
\[f,g\col\all\al.\Btree(\al)\a \Btree(\wild)\quad h\col
  \all\al.\Btree(\al)\a
  \Btree(\bot)\quad i\col\all\al.\Btree(\al)\a\Btree(\al) \]

The rewrite system with all type annotations is then

  \[
  \begin{array}{r@{~\a~}l}
    f\ \node(\al,\b)\ (\Node\ \al\ \b\ x\ y) &
    g\ \node(\al,\b)\ (i\ \node(\al,\b)\ (\Node\ \al\ \b\ x\ y)\\
    g\ \node(\al,\b)\ (\Node\ \al\ \b\ x\ y) & f\ \al\ (i\ \al\ x)\\
    g\ \leaf\ \Leaf & f\ \bot\ (h\ \leaf\ \Leaf)\\
    i\ \node(\al,\b)\ (\Node\ \al\ \b\ x\ y) &
    \Node\ \node(\al,\b)\ (i\ \al\ x)\ (i\ \b\ y)\\
    i\ \leaf\ \Leaf & \Leaf\\
    h\ \node(\al,\b)\ (\Node\ \al\ \b\ x\ y) & h\ \al\ x
  \end{array}
  \]

\section{Proof of theorem \ref{ho_dp:thm:sn}}
The proof uses computability
predicates (or candidates). 
As mentioned before, the absence of control, and particularly the
lack of orthogonality makes giving accurate semantics
somewhat difficult. We draw inspiration from the
termination semantics of Berger \cite{Berger05a}, which uses sets of
values to
denote terms. As is standard in computability proofs, each type will
be interpreted as a set of strongly normalizing (erased)
terms. Suppose a term $t$ reduces to the normal forms $\Leaf$ and
$\Node\ \Leaf\ \Leaf$. In that case $t$ is in the candidate that
contains all terms that reduce to $\Leaf$ or $\Node\ \Leaf\ \Leaf$, or
are hereditarily neutral. If $t$ the erasure of a term of type
$\Btree(\al)$ for some pattern variable $\al$, the interpretation
$\I{\Btree(\al)}$ must depend on some valuation of the free variable
$\al$. If we valuate $\al$ by some closed pattern $p$ and interpret
$\I{\Btree(\al)}$ by the set of terms whose normal forms are neutral
or match $p$, then the only possible choice for $p$ is
$\wild$. Clearly this does not give us the most precise possible
semantics for $t$, as it also includes terms such as $u=\Node\ (\Node\ x\
y)\ \Leaf$. However we need precise semantics if we are to capture the
information needed for the dependency analysis: if we take the
constructor term
$l=\Node\ Leaf\ x$, then a reduct of $t$ does match $l$, but this can
never happen for $u$. To give sufficiently precise semantics to terms,
we therefore need to interpret pattern variables with \emph{sets of
closed patterns}. In this case we will interpret $\al$ by the set
$\{\leaf,\node(\leaf,\leaf)\}$ to capture the most precise semantics
possible for $t$.


We define the interpretation of types, and prove that they satisfy the
Girard conditions. We then show that correctness of the defined
function symbols implies correctness of the semantics.

\begin{dfn}
  A \emph{value} is a term $v\in\efftrm$ of the form:
  
  \begin{itemize}
  \item $\lx.t$
  \item $\Node\ t\ u$
  \item $\Leaf$
  \end{itemize}
  For any $t\in\efftrm$ we say $v$ is a \emph{value of $t$} if
  $t\red^*_{\cR\cup\b}v$ and $v$ is a value.

  A term is \emph{neutral} if it is not a value, and is
  \emph{hereditarily neutral} if it has no values.
\end{dfn}
\bigskip

\begin{dfn}
  Let $\cP_{c}$ be the set of \emph{closed} patterns, and $\NF$ is the
  set of \emph{$\cR\b$-normal forms} in $\efftrm$.
  The \emph{term matching
    relation} $\redmatch\sle\NF\times\cP_{c}$
  is defined in the following way:
  \begin{itemize}
  \item $v\redmatch \_$
  \item $v\redmatch p$ if $v$ is neutral.
  \item $v\redmatch \node(p,q)$ if $v=
    \Node\ v_1\ v_2$ with $v_1\redmatch
    p \et v_2\redmatch q$.
  \item $v\redmatch \leaf$ if $v=\Leaf$.
  \end{itemize}

  A \emph{pattern valuation}, or \emph{valuation} if the context is
  clear, is a \emph{partial} function with finite support from pattern
  variables $\cV$ to \emph{non-empty} sets of \emph{closed}
  patterns. If $p$ is a
  pattern, $\t$ is a pattern valuation and $\FrV(p)\sle\dom(\t)$ then
  $p\t$ is the set defined inductively by:
  
  \begin{itemize}
  \item $\al\t   = \t(\al)$
  \item $\leaf\t = \leaf$
  \item $\wild\t = \wild$
  \item $\bot\t  = \bot$
  \item $\node(p_1,p_2)\t = \{\node(q_1,q_2)\mid q_1\in p_1\t\et
    q_2\in p_2\t\}$
  \end{itemize}
  We may write $p\t = \{p\mid
  \al_1\la\t(\al_1),\ldots,\al_n\la\t(\al_n)\}$, using inspiration
  from \emph{list comprehension} notation (as in Berger
  \cite{Berger05a}). If
  $\al\notin\dom(\t)$ and $P$ is a non-empty set of closed patterns,
  we write
  $\t\ext{al}{P}$ for the valuation that sends $\b\in\dom(\t)$ to
  $\t(\b)$ and $\al$ to $P$. Notice that $p\t$ is a set of
  \emph{closed} patterns.
  
  Finally if $\t$ is a valuation and $t$ is a term in $\SN$, we write
  $t\redmatch p\t$ if for
  every normal form $v$ of $t$:
  \[\exists q\in p\t,\ v\redmatch q \]

  \bigskip

  The \emph{type interpretation} $\I{\_}_{\_}$ is a function that to each
  $T\in\cT$ and each valuation $\t$ such that $\FrV(T)\sle\dom(\t)$
  associates a
  set $\I{T}_\t\sle\SN_{\cR\cup\b}$. We define it by induction on the
  structure of $T$:
  
  \begin{itemize}
  \item $\I{\Btree(p)}_\t= \{t\in\cB\mid t\redmatch p\t\}$
  \item $\I{T\a U}_\t= \{t\in\SN\mid\all u\in \I{T}_\t,
    t\ u\in\I{U}_\t\}$
  \item $\I{\all\al.T}_\t=\{t\in\SN\mid\all P, t\in
  \I{T}_{\t\ext{\al}{P}}\}$
  \end{itemize}

  Where $\cB$ is the smallest set that verifies:
  \[\cB=\{t\in\SN\mid \all v\mbox{ a value of }t, v=\Leaf\ou
  v=\Node\ t_1\ t_2\et t_1,t_2\in\cB\} \]
\end{dfn}

The next step in the reducibility proof is to verify that the
interpretation of terms verify the \emph{Girard conditions}: A subset
$X\sle\efftrm$ satisfies the Girard conditions if

\begin{enumerate}
\item strong normalization: $X\sle\SN$
\item stability by reduction: for every term $t\in X$, if $u$ is such
  that $t\red^* u$, then $u\in X$.
\item ``sheaf condition'': if $t$ is \emph{neutral}, and for every
  term $u$ such that $t\red u$, $u\in X$, then $t\in X$.
\end{enumerate}

These embody the exact combinatorial properties required to carry
through the inductive proof of correctness, namely that every
well-typed term is in the interpretation of its type.

\begin{lem}\label{dp_prf:lem:girard_cond}
  If $T\in\cT$ is a type, then for every valuation
  $\t$,
  \[\I{T}_\t\mbox{ satisfies the girard conditions} \]
\end{lem}

\begin{prf}
  We proceed by induction on the structure of $T$.
  
  \begin{itemize}
  \item $T=\Btree(p)$
    
    \begin{itemize}
    \item Strong normalization: by definition of $\cB$.
    \item Stability by reduction. Suppose that
      $t\in\I{\Btree(p)}_\t$. If $t\red^* u$, then the set of normal
      forms of $u$ is contained in the set of normal forms of $t$.
    \item Sheaf condition. Suppose that $t$ is neutral and that
      each one step reduct of $t$ is in $\I{\Btree(p)}_\t$. Now either
      $t$ is in normal form, and then $t\redmatch  p\t$ (as it is non
      empty), or for every one step reduct $u$ of $t$, $u\redmatch
      p\t$. But in this case every normal form of $t$ is the normal
      form of some $t\red u$, and thus $t\redmatch p\t$.

    \end{itemize}
    
  \item $T=T_1\a T_2$ 
    
    \begin{itemize}
    \item Strong normalization: by definition.
    \item Stability by reduction. Simple application of induction
      hypothesis.
    \item Sheaf condition: Let $t$ be neutral and suppose that
      $t'\in\I{T\a U}_\t$ for every $t'$ a reduct of $t$. Let $u$ be
      an arbitrary element of $\I{T}_\t$. Then $t'\ u'\in\I{U}_\t$ for
      every reduct $u\red^*u'$, by
      definition of the interpretation and stability by reduction. By
      induction hypothesis, this implies that $t\ u\in\I{U}_\t$, as it
      is again a neutral term. As $u$ was chosen arbitrarily, then
      $t$ is in $\I{T\a U}_\t$.
    \end{itemize}
    
  \item $T=\all\al.U$
    
    \begin{itemize}
    \item Strong normalization: by definition.
    \item Stability by reduction: Let $t\in\I{\all\al.U}_\t$. We have
      for every set $P$ of closed terms, $t\in\I{U}_{\t\ext{\al}{P}}$.
      By induction, every reduct $u$ of $t$ is also in
      $\I{U}_{\t\ext{\al}{P}}$. As $P$ was chosen arbitrarily, $u$ is
      also in $\I{\all\al.U}_\t$.
    \item Sheaf condition. Let $t$ be neutral and suppose that one
      step reducts of $t$ are in $\I{\all\al.U}_\t$. Take an arbitrary
      $P$. Every reduct of $t$ is in $\I{U}_{\t\ext{\al}{P}}$. By
      induction hypothesis, $t$ is in $\I{U}_{\t\ext{\al}{P}}$, from
      which we may conclude.
    \end{itemize}
    
  \end{itemize}
  \cqfd
\end{prf}

Now we give the conditional correctness theorem, which states that if
the function symbols belong to the interpretation of their types, then
so does every well-typed term.

\begin{dfn}
  Let $\t$ be a pattern valuation, $\s$ a substitution
  from term variables to erased terms, and $\G$ a context. We say that
  $(\t,\s)$ \emph{validates} $\G$,
  and we write $\s\models_\t\G$, if the set of free pattern variables
  in $\G$
  is contained in $\dom(\t)$, and if for every $x\in\dom(\G)$
  \[ \s(x)\in\I{\G x}_\t\]
  Likewise, we write $\s\models_\t t\col T$ if
  $\FrV(t)\sle\dom(\s)$, $\FrV(T)\sle\dom(\t)$ and 
  \[ \erase{t}\s\in\I{T}_\t\]
\end{dfn}

\begin{thm}\label{dp_prf:thm:rel_correct}
  Suppose that for each $f\in\S$ and each valuation
  $\t$,
  \[ f\in\I{\tau_f}_\t \]
  then for every context $\G$, term $t$ and type $T$, if $\G\th t\col
  T$
  \[\all (\t,\s),\ \s\models_\t\G\A \s\models_\t t\col T \]
\end{thm}

We need the classic \emph{substitution lemma} for types:

\begin{lem}\label{dp_prf:lem:subst_pat}
  For every patterns $q,p$ and valuation $\t$, if $\al$ is not in the
  domain of $\t$, then
  \[p\subst{\al}{q}\t=p\t\ext{\al}{q\t} \]
\end{lem}

\begin{prf}
  We proceed by induction on the structure of $p$:
  
  \begin{itemize}
  \item $p=\al$: trivial.
  \item $p=\b\neq\al$: We have $p\subst{\al}{q}=\b$ and therefore
    $p\subst{\al}{q}\t=\t(\b)=\t\ext{\al}{q\t}(\b)$.
  \item $p=\leaf,\wild,\bot$: trivial.
  \item $p=\node(p_1,p_2)$: We have
    $p\subst{\al}{q}\t=\node(p_1\subst{\al}{q},p_2\subst{\al}{q})\t$. But
    this last term is equal to
    \[\{\node(q_1,q_2)\mid q_i\in p_i\subst{\al}{q}\t,i=1,2\}\]
    which by induction is equal to
    \[\{\node(q_1,q_2)\mid q_i\in p_i\t\ext{\al}{q\t},i=1,2\} \]
    which allows us to conclude.
  \end{itemize}
  \cqfd
\end{prf}

\begin{lem}{\bf (substitution lemma)}\label{dp_prf:lem:subst}

  Let $T$ be a type and $\t$ a pattern valuation. If $\al$
  does not appear in the domain of $\t$
  then:
  \[ \I{T\subst{\al}{p}}_\t=\I{T}_{\t\ext{\al}{p\t}} \]
\end{lem}

\begin{prf}
  We proceed by induction on the type.
  \begin{itemize}
  \item Atomic case:
    \[\I{\bs{B}(q)\subst{\al}{p}}_\t=\left\{t\in\SN\mid
    t\redmatch
    q\subst{\al}{p}\t\right\}\] 
    But by lemma \ref{dp_prf:lem:subst_pat},
    $q\subst{\al}{p}\t=p\t\ext{\al}{p\t}$, from which we can conclude.
  \item Arrow case: straightforward from induction hypothesis.
  \item case $\all \b.T$. We may suppose by Barendregts convention that
    $\b$ is distinct from $\al$, not in the domain of $\t$
    and distinct from all variables in $p$. We then have:
    \[ \I{(\all \b.T)\subst{\al}{p}}_\t= \{t\in\SN\mid \all
    Q,t\in\I{T\subst{\al}{p}}_{\t\ext{\b}{Q}}\}\]
    Let $\t'=\t\ext{\b}{Q}$. We may apply the induction hypothesis, which
    gives:
    \[ \I{T\subst{\al}{p}}_{\t'} = \I{T}_{{\t'}\ext{\al}{p\t'}}\]
    And as $\b$ does not appear in $p$:
    \[ \I{T}_{{\t'}\ext{\al}{p\t'}}=\I{T}_{\t\ext{\al\ \ \b}{p\t\ Q}}\]
    But we have:
    \[ \{t\in\SN\mid\all Q, t\in \I{T}_{\t\ext{\al\ \ \b}{p\t\
        Q}}\}=\I{\all \b.T}_{\t\ext{\al}{p\t}}\]
    Which concludes the argument.
  \end{itemize}\cqfd
\end{prf}

We may easily generalize this result to:

\begin{cor}\label{dp_prf:cor:subst}
  Let $T$ be a type. If $\phi$ is a substitution, and $\t$ is a
  valuation such that the variables of $T$ do not appear in
  the domain of $\t$, then:
  \[ \I{T\phi}_\t=\I{T}_{\t\circ\phi}\]
  Where $\t\circ\phi$ is the valuation defined by $\t\circ\phi(\al)
  = \phi(\al)\t$.
\end{cor}

Another useful lemma states that type interpretations only depend on
the value of the pattern substitutions in the free variables of the
type.

\begin{lem}\label{dp_prf:lem:eq_subst}
  Let $T$ be some type and $\t,\t'$ be two closed pattern
  substitutions. If $\t(\al)=\t'(\al)$ for every $\al\in\FrV(T)$, then
  $\I{T}_\t=\I{T}_{\t'}$.
\end{lem}
\begin{prf}
  Straightforward induction on $T$.
\end{prf}

The next lemmas show correctness of the interpretation with
respect to subtyping.

\begin{dfn}
  Let $P$ and $Q$ be sets of closed patterns. We write $P\subpat Q$ if
  for each $p\in P$, there is a $q\in Q$ such that $p\subpat q$.
\end{dfn}

\begin{lem}\label{dp_prf:lem:sub_subst}
  Let $\t$ be a pattern valuation. If $p\subpat q$, then $p\t\subpat q\t$
\end{lem}

\begin{prf}
  Induction on the derivation of $p\subpat q$. The only interesting
  case is $\node(p_1,p_2)\subpat\node(q_1,q_2)$ with $p_i\subpat q_i$
  for $i=1,2$. In that case, if $r\in\node(p_1,p_2)\t$, we have
  $r=\node(r_1,r_2)$ with $r_i\in p_i\t$ for $i=1,2$. By induction
  hypothesis, there is $r'_1, r'_2$ in $\node(q_1,q_2)\t$ such that
  $r_i\subpat r'_i$ for each $i$. Then we take
  $\node(r'_1,r'_2)\in\node(q_1,q_2)\t$ to conclude.

  \cqfd
\end{prf}

\begin{lem}\label{dp_prf:lem:correct_sub}
  Suppose $T\leq U$. Then for all $\t$, $\I{T}_\t\sle\I{U}_\t$
\end{lem}

\begin{prf}
  We proceed by induction on all the possible cases for the judgement
  $T\leq U$.
  \begin{itemize}
  \item $p\subpat q$: We first show that for all terms $t$, and every
    non-empty set of closed patterns $P$ and $Q$, if $P\subpat Q$,
    then $t\redmatch P\A t\redmatch Q$. This follows from the
    following fact: if $v$ is in normal form and $r\subpat s$, then
    \[v\redmatch r \A v\redmatch s\]    
    To show this we proceed by induction on the
    $\subpat$ judgement. The first three cases are easy. In the fourth
    case, $v\redmatch \node(r_1,r_2)$ which by definition implies that
    $v=\Node\ v_1\ v_2$, with $v_1\redmatch r_1$ and
    $v_2\redmatch r_2$. We can then conclude by the induction hypothesis.

    Now using lemma \ref{dp_prf:lem:sub_subst}, we have, if $p\subpat
    q$, $t\redmatch p\t\A t\redmatch q\t$.
    \bigskip

    Now let $t\in\I{\Btree(p)}_\t$, we have by definition $t\redmatch
    p\t$, and by the previous remark, $t\redmatch q\t$ which implies
    $t\in\I{\Btree(q)}_\t$.
  \item Suppose $T_2\leq T_1$ and $U_1\leq U_2$. Let $t$ be in
    $\I{T_1\a U_2}_\t$, we show that it is in $\I{T_2\a U_2}_\t$. Let
    $u$ be in $\I{T_2}_\t$. By the induction hypothesis,
    $u\in\I{T_1}_\t$, therefore (by definition of $\I{T_1\a U_1}_\t$),
    $t\ u$ is in $\I{U_1}_\t$, which by another application of the
    induction hypothesis, is included in $\I{U_2}_\t$. From this we
    can conclude that $t$ is in $\I{T_2\a U_2}_\t$.
  \item Let $t$ be a term in $\I{\all\al.T}_\t$ and $P$ be some
    arbitrary set of closed patterns, and suppose that $\al$ is a
    variable
    not appearing in the domain of $\t$. We then have
    $$ t \in \I{T}_{\t\ext{\al}{P}}$$
    Since $\all\al.T\leq \all\al.U$, we have $T\leq
    U$. The induction hypothesis gives:
    $$\I{T}_{\t'}\sle \I{U}_{\t'} $$
    for all valuations $\t'$. Take $\t'$ to be
    $\t\ext{\al}{P}$. We have:
    $$ \I{T}_{\t\ext{\al}{P}}\sle \I{U}_{\t\ext{\al}{P}} $$
    From this we can deduce $t\in \I{U}_{\t\ext{\al}{P}}$ and conclude.
  \end{itemize}\cqfd
\end{prf}

We shall also need the fact that given $T$ and a valuation $\t$, then
$\I{T}_\t$ is included in $\I{T}_{\t'}$ if $\t'$ is a \emph{weakening}
of $\t$ on the variables in \emph{positive position} in $T$.

\begin{lem}\label{dp_prf:lem:leq_subst}
  Let $T$ be a type and $\t,\t'$ two pattern valuations.
  If $\t(\al)\subpat\t'(\al)$ for every free variable $\al\in T$ in a
  \emph{positive} position, and $\t(\b)=\t'(\b)$ for every other
  variable, then 
  \[\I{T}_\t\sle\I{T}_{\t'}\]
  Conversely if
  $\t(\al)\subpat\t'(\al)$ for every free variable $\al$ in a
  \emph{negative} position, then
  \[\I{T}_{\t'}\sle\I{T}_\t \]
\end{lem}

\begin{prf}
  First notice that if $p$ is a pattern, then $p\t\subpat p\t'$, by a
  simple induction on $p$. We prove both propositions simultaneously
  by induction on $T$:
  
  \begin{itemize}
  \item $T=\Btree(p)$. All variables of $p$ appear positively in
    $T$. Then by the above remark, $p\t\subpat p\t'$,
    and therefore $\I{\Btree(p)}_\t\sle\I{\Btree(p)}_{\t'}$.
  \item $T=T_1\a T_2$. We treat the positive case. We have by
    induction hypothesis
    $\I{T_1}_{\t'}\sle\I{T_1}_{\t}$, as all variable of $T_1$ that
    appear positively in $T$ appear negatively in $T_1$, and
    $\I{T_2}_{\t}\sle\I{T_2}_{\t'}$. Therefore, by definition of
    $\I{T_1\a T_2}_\phi$, we have:
    \[\I{T_1\a T_2}_\t\sle\I{T_1\a T_2}_{\t'} \]

    The negative case is treated in the same fashion.
  \item $T=\all\al.U$: straightforward induction.
  \end{itemize}

  \cqfd
\end{prf}
\bigskip

We can now prove the correctness of the interpretation relative to
that of the function symbols
(theorem \ref{dp_prf:thm:rel_correct}).

\begin{prf}
  We proceed by induction on the typing derivation.
  \begin{itemize}
  \item ax: by definition of $\s\models_\t\G$.
  \item t-lam: By induction hypothesis, for all $\s',\t'$ such that
    $\s'\models_{\t'}\G,x\col T$, $t\s'$ is in $\I{U}_{\t'}$. By
    definition of $\I{T\a U}_\t$, we need to show that for any
    $u\in\I{T}_\t$, $(\lx\col T.t)\s u$ is in $\I{U}_{\t}$. Now as
    this term is neutral, it suffices to show that every reduct is in
    $\I{U}_{\t}$. We proceed by well founded induction on the reducts
    of $t$ and $u$. Thus if $(\lx\col T.t)\s u\red (\lx\col T.t')\s
    u'$ with $t\red t'$ or $u\red u'$, then we may conclude by
    well-founded induction hypothesis. The remaining case is $(\lx\col
    T.t)\s u\red t\s\subst{x}{u}$. To show that this is in
    $\I{U}_{\t}$, we apply the main induction hypothesis with
    $\s'=\s\ext{x}{u}$ and $\t'=\t$.
    
    It can be argued that this argument is the fundamental combinatory
    explanation for normalization of $\b$-reduction.
  \item p-lam: by induction hypothesis, for all $\s',\t'$ such that
    $\s'\models_{\t'}\G,
    \erase{t}\s'$ is in $\I{T}_{\t'}$. Let $\s,\t$ be some such
    valuations and $P$ be a set of closed patterns. As
    $\erase{\lam\al.t}\s=\erase{t}\s$,
    we need to show that $\erase{t}\s\in\I{T}_{\t\ext{\al}{P}}$

    Observe that if $\al$ does not appear in $\G$, then
    $\s\models_\t\G$ implies $\s\models_{\t\ext{\al}{P}}\G$, by virtue of
    lemma \ref{dp_prf:lem:eq_subst}. We may therefore conclude that
    $\erase{t}\s$ is in $\I{T}_{\t\ext{\al}{P}}$.

  \item leaf-intro: Clear by definition of $\I{\Btree(\leaf)}_\t$
    
  \item node-intro: let $t,u$
    be terms in $\I{\Btree(\al)}_\t$ and
    $\I{\Btree(\b)}_\t$, respectively. The normal forms of
    $\Node\ t\ u$ are of the form $\Node\ t'\ u'$,
    with $t'$ and $u'$ normal forms of $t$ and $u$,
    respectively. Therefore, to check if
    $\Node\ t\ u\redmatch\node(\t(\al),\t(\b))$, it suffices to check
    $t\redmatch\t(\al)$ and $u\redmatch\t(\b)$, both of which are true
    by hypothesis.

  \item t-app: straightforward by the induction hypothesis.
  \item p-app: by hypothesis, $\erase{t}\s\in\I{\all x.T}_\t$, this gives by
    definition $\erase{t}\s\in\I{T}_{\t\ext{x}{p\t}}$, and by the
    substitution lemma
    (lemma \ref{dp_prf:lem:subst}),
    $\erase{t}\s\in\I{T\subst{x}{p}}_\t$, therefore
    \[\erase{t\ p}\s\in\I{T\subst{x}{p}}_\t\]
  \item symb: By hypothesis.
  \item sub: By application of the correctness of subtyping (lemma
    \ref{dp_prf:lem:correct_sub}), and the induction hypothesis.
  \end{itemize}\cqfd
\end{prf}


Now it remains to show that each function symbol is computable. By
analogy with the first-order dependency pair framework, we need to
build an order on terms that is in relation to the approximated
dependency graph. Then sequences of decreasing terms will be the
analogue of \emph{chains}, and we will show that there can be no
infinite decreasing
sequences. Instead of actual terms, it is more convenient, when
dealing with higher-order rewriting, to order tuples of terms labeled
by a head function symbol, \ie instead of having $f\vt>g\vu$ we have
$(f,\vt)>(g,\vu)$. The reason for this is that recursive calls in the
right-hand side of rewrite rules needn't be applied to all their
arguments. We will therefore need a way of using typing to ``predict''
which arguments may be applied, using the order on tuples as above.
\bigskip

However it is quite subtle to build this order in practice: indeed, a
natural candidate for such an order is (the transitive closure of) the
order defined by $(f,\vt)>(g,\vu)$ if and only if 
\[ \ex\t,\phi, f^\sharp(p_1,\ldots,p_n)\red
g^\sharp(q_1,\ldots,q_m)\in\cG,\ \all i,j,\ t_i\in\I{\Btree(p_i)}_\t\et
u_j\in\I{\Btree(q_j)}_\phi \]
This would allow us to easily build the relation between the graph and
the order, and show that each call induces a decrease in this
order. Sadly, this order may not be well founded even in the event
that the termination criterion is satisfied. Consider for example the
rule $f\ \node(\al,\b)\ (\Node\ x\ y)\red f\ \al\ x$, typeable in the
context $\G=x\col\Btree(\al),y\col\Btree(\b)$. Given the above
definition, we have $(f,t)>(f,u)$ provided that there are closed $p$
and $q$ such that $t\redmatch p$ and $u\redmatch q$. But then we may
take $p=q=\wild$ and if $t=z$ and $u=z$ with $z$
a variable, then $(f,z)>(f,z)$. The rewrite system does satisfy the
criterion, as $\node(\al,\b)\tgt\al$, but the order is not well
founded.

One possible solution is to restrict the reduction to call-by value on
closed terms, where a reduction in $\cR$ can occur only if the
arguments to the defined function are in normal form, and values
(although $\b$-reduction can occur at any moment). However we
strive for more generality. 

Another solution, in the previous example, is to impose the
condition that $t$ must be equal to $\Node\ t_1\ t_2$, which makes the
counter-example invalid. However, we still do not have any necessary
relationship between $t$ and $u$, and we may take in particular
$t=\Node\ x\ y$ and $u=\Node\ x\ y$, which again results in a non well
founded sequence. The solution is to take, instead of just a particular
instance of the pattern variables, the \emph{most general} possible
instance.

\begin{dfn}
  Take the set $\cP_{min}$ of \emph{minimal patterns} to be the subset
  of $\cP$ defined by:
  \[p,q\in\cP_{min}\ceq \al\mid \leaf\mid \node(p,q) \]

  Let $t$ be a term in normal form. We inductively define the
  \emph{pattern form} $pat(t)$ of $t$ inductively:
  
  \begin{itemize}
  \item $pat(t)=\bot$ if $t$ is neutral.
  \item $pat(\Leaf)=\leaf$
  \item $pat(\Node\ t\ u)=\node(pat(t),pat(u))$
  \item $pat(t)=\wild$ otherwise.
  \end{itemize}
  
  We define the partial \emph{type matching}
  function $\match_\cP$ that takes terms $t_1,\ldots,t_n$ in
  $\efftrm$, and minimal patterns
  $p_1,\ldots,p_n$ in $\cP_{min}$ and returns a pattern valuation:
  
  \begin{itemize}
  \item if $p_1=\al_1,\ldots,p_n=\al_n$ and $t_i=t_j$ whenever
    $\al_i=\al_j$, then 
    \[\match_\cP(\vt,\vp)(\al_i)=t_i\]
  \item if $p_i=\node(q_1,q_2)$ and $t_i=\Node\ u_1\ u_2$ then
    \[\match_\cP(\vt;\vp) =
    \match_\cP(t_1,\ldots,t_{i-1},u_1,u_2,t_{i+1},\ldots,t_n
    \ ;\   p_1,\ldots,p_{i-1},q_1,q_2,p_{i+1},\ldots,p_n) \]
  \item if $p_i=\leaf$ and $t_i=\Leaf$ then
    \[\match_\cP(\vp,\vt)=\match_\cP(t_1,\ldots,t_{i-1},t_{i+1},\ldots,t_n  
    \ ; \  p_1,\ldots,p_{i-1},p_{i+1},\ldots,p_n)\]
  \item $\match_\cP$ is undefined in other cases.
  \end{itemize}
  
\end{dfn}

The type matching can be seen as a way of giving the most precise
possible valuation for terms that match some
left-hand side of a rule. Notice that for each $f^\sharp(\vp)\red
g^\sharp(\vq)\in\cG$, each $p_i$ is in $\cP_{min}$. Indeed, an
examination of the minimal typing rules show that only minimal
patterns may appear in types.

Note also that if $\match(\vt,\vp)=\t$, then for each $i$,
$t_i\redmatch p_i\t$, by a simple induction. 

\begin{dfn}
  A \emph{link} is a tuple $(n,\vt,\vu)$ such that
  
  \begin{itemize}
  \item $\vt,\vu\in\SN$
  \item $n=f^\sharp(\vp)\red g^\sharp(\vq)\in\cG$
  \item $\match_\cP(\vt,\vp)$ is defined, and if it is equal to $\t$,
    then
    \[\all j, u_j\redmatch q_j\t' \]
    For some extension $\t'$ of $\t$ such that
    $\FrV(\vq)\sle\dom(\t')$.
  \end{itemize}

  A \emph{chain} is an eventually infinite sequence $c_1,c_2,\ldots$
  of links such that if $c_i=(n_i,\vt_i,\vu_i)$, then for each $i$,
  \[\vu_i\red^*\vt_{i+1} \]
  and if $n_i=f_i^\sharp(\vp)\red g_i^\sharp(\vq)$ then $g_i=f_{i+1}$.
  
\end{dfn}
Notice that if $\FrV(\vq)\sle\FrV(\vp)$ then we may take $\t'=\t$ in
the definition of chains.
\bigskip

We first need to show a correspondence between the chains and the
graph, that is:

\begin{lem}\label{dp_prf:lem:correct_graph}
  For each chain $c_1,c_2,\ldots$ such that $c_i=(n_i,\vt_i,\vu_i)$,
  there is a path $n_1\a n_2\a\ldots$ in $\cG_\cR$.
\end{lem}

\begin{prf}
  It suffices to show that if
  $c_1=(n_1,\vt,\vu),c_2=(n_2,\vu',\vv)$ is a chain, then there
  is an
  edge between $n_1=f^\sharp(\vp)\red g^\sharp(\vq)$ and
  $n_2=g^\sharp(\vr)\red h^\sharp(\vs)$. First note that the variables
  of $\vq$ and $\vr$ are distinct by hypothesis. Notice that for each $i$,
  $u_i\redmatch q_i\t$ for some $t$ and $\match_\cP(r_i,u'_i)$ is
  defined. We need to prove for each $i$
  that $q_i\bowtie r_i$. As $u_i\red^* u'_i$, all normal forms of
  $u'_i$ are also normal forms of $u_i$. We proceed by induction on
  $\match_\cP(r_i,u_i')$.
  
  \begin{itemize}
  \item $r_i$ is a variable. We can conclude immediately by the
    definition of $\bowtie$, as a fresh variable can unify with any
    pattern.
  \item $u'_i=\Leaf$ and $r_i=\leaf$. In this case $\Leaf$ is a normal
    form of $u_i$,
    so there is some $q'\in q_i\t$ such that $\Leaf\redmatch q'$. From
    this it follows that $q_i$ is either $\leaf$, $\wild$ or some
    variable. This allows us to conclude that $q_i\bowtie\leaf$.
  \item $u'_i=\Node\ u'^1_i\ u'^2_i$ and $r_i=\node(r_i^1,r_i^2)$. Now
    let us examine $q_i$. We may exclude the cases $q_i=\leaf$ and
    $q_i=\bot$, as every normal form of $u'_i$ is a normal form of
    $u_i$ and is of the form $\Node\ v\ v'$. In the case $q_i=\al$ or
    $q_i=\wild$ we may easily conclude. The only remaining case is
    $q_i=\node(q_i^1,q_i^2)$. From the induction hypothesis we get
    $q_i^1\bowtie r_i^1$ and $q_i^2\bowtie r_i^2$, which imply
    $q_i\bowtie r_i$
  \end{itemize}
  
  \cqfd
\end{prf}

If the conditions of the termination theorem are satisfied, the there
are no infinite chains, in the same way as for the first-order
dependency pair approach.

\begin{thm}\label{dp_prf:thm:finite_chains}
  Suppose that the conditions of theorem \ref{ho_dp:thm:sn} are
  satisfied. Then there are no infinite chains.
\end{thm}

We need to define and establish the well foundedness of the embedding
order on terms.

\begin{dfn}
  We mutually define the strict and large \emph{embedding preorder} on
  erased terms in normal form $\tgt$ and $\tge$ by:
  
  \begin{itemize}
  \item $t_1\tge u\A \Node\ t_1\ t_2\tgt u$
  \item $t_2\tge u\A \Node\ t_1\ t_2\tgt u$
  \item $t_1\tgt u_1\et t_2\tge u_2\A \Node\ t_1\ t_2\tgt
    \Node\ u_1\ u_2$
  \item $t_1\tge u_1\et t_2\tgt u_2\A \Node\ t_1\ t_2\tgt
    \Node\ u_1\ u_2$
  \item $\Leaf\tge \Leaf$
  \item $t\tge u$ if $t$ and $u$ are neutral.
  \item $t\tgt u \A t\tge u$
  \end{itemize}

  Note that the preorder is \emph{not} an order: for instance, $x\tge
  y$ and $y\tge x$.
\end{dfn}

\begin{lem}\label{dp_prf:lem:tgt_wf}
  The preorder $\tgt$ is well-founded.
\end{lem}

\begin{prf}
  Given a term in normal form $t$, define $\size(t)$ inductively:
  
  \begin{itemize}
  \item $\size(\Node\ t_1\ t_2)=\size(t_1)+\size(t_2)+1$
  \item $\size(t)=0$ otherwise.
  \end{itemize}
  It is then easy to verify by mutual induction that if $t\tgt u$,
  $\size(t)>\size(u)$ and if $t\tge u$ then
  $\size(t)\geq\size(u)$. Well foundedness of the order on naturals
  yields the desired conclusion.
  
  \cqfd
\end{prf}

To show that there are no infinite chains, we will exploit the fact
that if $c=(n,\vt,\vu)$ is a link, that is decreasing
in the embedding order on patterns, then there is a decrease in the
normal forms from $t$ to $u$.

To show this, we must prove that pattern-matching does indeed
\emph{completely} capture the ``pattern semantics'' of a term in
$\cB$.

\begin{lem}\label{dp_prf:lem:trm_to_pat}
  Suppose $\vt$ are terms in $\cB$ and $\vp$ are minimal patterns. If
  $\match_\cP(\vt,\vp)$ is defined and equal to $\t$, then for each
  $q\in p_i\t$, there is a normal form $v$ of $t_i$ such that
  $pat(v)=q$.
\end{lem}

\begin{prf}
  We proceed by induction on the definition of $\match_\cP$:
  
  \begin{itemize}
  \item $p_i=\al_i$. In this case (as $\match_\cP(\vt,\vp)$ is defined)
    By definition $\al\t$ is equal to $\{pat(v)\mid v\mbox{ is a
      normal form of }t_i\}$.
  \item $p_i=\leaf$. In this case $t_i=\Leaf$ and therefore we can
    take $v=\Leaf$.
  \item $p_i=\node(p_i^1,p_i^2)$. In this case,
    $t_i=\Node\ t^1_i\ t^2_i$. By the induction hypothesis, for any
    $q_1\in p_i^1\t$ and $q_2\in p_i^2\t$ there are
    normal forms $v_1$ and $v_2$ of $t_i^1$ and $t_i^2$ such that
    $q_j=pat(v_j)$ for $j=1,2$. It is easy to observe that
    $\Node\ v_1\ v_2$ is a normal form of $t_i$, and that
    $q=\node(q_1,q_2)$ is an element of $p_i\t$, and
    $pat(\Node\ v_1\ v_2)=q$ allows us to conclude.
  \end{itemize}
  
  \cqfd
\end{prf}

To prove that there are no infinite chains, we need to relate the
decrease of the patterns to the decrease of the normal forms of the
terms that appear in chains.

\begin{lem}\label{dp_prf:lem:pat_emb_trm_emb}
  Suppose that $p$ and $q$ are closed patterns such that $p\tgt q$
  (respectively $p\tge q$), and $v_1,v_2$
  normal forms such that $pat(v_1) = p$ and $v_2\redmatch q$. Then 
  $v_1\tgt v_2$ (respectively $v_1\tge v_2$).
\end{lem}

\begin{prf}
  We prove both properties simultaneously by induction on the
  derivation of $p\tgt q$:
  
  \begin{itemize}
  \item $p=\node(p_1,p_2)$ and $p_1\tge q$. We have
    $v_1=\Node\ u_1\ u_2$ with $pat(u_1)=p_1$. By induction hypothesis
    $u_1\tge v_2$, and therefore $\Node\ u_1\ u_2\tgt v_2$.
  \item $p=\node(p_1,p_2), q=\node(q_1,q_2)$ with $p_1\tgt q_1$ and
    $p_2\tge q_2$. In that case $v_1=\Node\ v_1^1\ v_1^2$ and
    $v_2=\Node\ v_2^1\ v_2^2$. The induction hypothesis gives
    $v_1^1\tgt v_2^1$ and $v_1^2\tge v_2^2$, from which we may conclude.
  \item The symmetrical cases are treated in the same manner.
  \item $p=\leaf$ and $q=\leaf$. In this case, $v_1=v_2=\Leaf$, and
    $v_1\tge v_2$.
  \end{itemize}
  
\end{prf}

\begin{lem}\label{dp_prf:lem:chain_decr}
  Let $c=(n,\vt,\vu)$ be some link such that $n=f^\sharp(\vp)\red
  g^\sharp(\vq)$. Suppose that there is $i$ such that $p_i\tgt q_i$,
  (respectively $p_i\tge
  q_i$). Then if $v$ is a normal form of
  $u$, there exists some normal form $v'$ of $t$ such that $v'\tgt v$,
  (respectively $v'\tge v$).
\end{lem}

\begin{prf}
  Let $\t=\match_\cP(\vt,\vp)$, which is guaranteed to exist by
  hypothesis. First notice that for every $\al\in\FrV(\vp)$, $\t(\al)$
  does \emph{not} contain $\_$. Indeed, given $t\in\cB$, the normal
  form of $t$ is also in $\cB$. It can only be neutral, equal to
  $\Leaf$, or in the form $\Node\ t_1\ t_2$ with $t_i$ in the above
  form.

  We treat the $\tgt$ case first.
  Suppose that $v$ is a normal form of $u_i$. By definition, we have
  $u_i\redmatch q_i\t$, which means by definition that there is some
  $r\in q_i\t$ such that $v\redmatch r$. Since $p_i\tgt q_i$, this
  implies that there is
  some $r'\in p_i$ such that $r'\tgt r$. We have by lemma
  \ref{dp_prf:lem:trm_to_pat} that there exists some $v'$ a normal
  form of $t_i$ such that $pat(v')=r'$, which allows us to conclude
  using lemma \ref{dp_prf:lem:pat_emb_trm_emb}.

  \cqfd
\end{prf}

We finally have all the tools to give the proof of well foundedness of
chains.

\begin{prf} of theorem \ref{dp_prf:thm:finite_chains}.

  By contradiction, let $c_1,c_2,\ldots$ be an infinite chain, such
  that for each $i$, $c_i=(n_i,\vt_i,\vu_i)$. By
  lemma \ref{dp_prf:lem:correct_graph}, $n_1,n_2,\ldots$ is an
  infinite path in $\cG$. By finiteness of $\cG$, there is some SCC
  $\cG'$
  and some natural number $k$ such that $n_k,n_{k+1},\ldots$ is
  contained in $\cG'$. By hypothesis, if $n_i=f_i^\sharp(\vp^i)\red
  g_i^\sharp(\vq^i)$, there is an index $j$ such that for each $i$,
  $p^i_j\tge q^i_j$ or $p^i_j\tgt p_i^j$. Furthermore, again by
  hypothesis, there are an
  infinite number of indexes $i$ such that $p^i_j\tgt q^i_j$. Let
  $V_i=\{v\mid v\mbox{ is a normal form of }t_j^i\}$ and $U_i= \{v\mid
  v\mbox{ is a normal form of }u_j^i\}$. We apply
  lemma \ref{dp_prf:lem:chain_decr} to show that for each $v'_i\in
  U_i$ there exists $v_i\in V_i$ such that $v_i\tgt v'_i$ for these
  indexes and $v_i\tge v'_i$ for the others.

  We wish to show that there is an infinite chain $v_1,v_2,\ldots$
  such that $v_i\tge v_{i+1}$ for each $i$ and $v_i\tgt v_{i+1}$ for
  an infinite number of indexes $i$, contradicting well-foundedness of
  $\tgt$ (lemma \ref{dp_prf:lem:tgt_wf}).

  To do this we first notice that $V_{i+1}\sle U_i$, as $\vu_i\red^*
  \vt_{i+1}$. Then we build the following tree:
  
  \begin{itemize}
  \item We have a node at the top, connected to every element of
    $V_k$.
  \item We have a node between $a\in V_i$ and $b$ in $U_i$ if $a\tge
    b$ or $a\tgt b$.
  \item We have a node between $a\in U_i$ and $b\in V_{i+1}$ if $a=b$.
  \end{itemize}
  Notice first that every $V_i,U_i$ is finite, as the rewrite system
  is finite (each strongly normalizing term therefore has a finite
  number of normal forms). We wish to apply \emph{K\"onig's lemma}
  which states: every finitely branching infinite tree has an infinite
  path. It is easy to see that the tree is finitely branching: every
  $V_i$ and $U_i$ is finite, and it is equally easy to verify that the
  tree is infinite, as no $V_i$ or $U_i$ is empty (the $t_i$ and $u_i$
  are strongly normalizing and therefore have at least one normal
  form). This give us the existence of an infinite
  path in the tree, which concludes the proof.

  \cqfd
\end{prf}

\bigskip

To prove that the function symbols are in the interpretation of their
type, we shall (obviously) need to consider the rewrite rules. In
particular, we need to relate the minimal typing used to derive the
types of left hand sides and pattern matching, in order to prove that
our notion of chain is the correct one.

\begin{lem}\label{dp_prf:lem:min_typ_match}
  Suppose that $\G$ is a context, that $l_1,\ldots, l_k$ are
  constructor terms and that
  $\G\th_{min}l_1\col\Btree(p_1),\ldots,\G\th_{min}l_k\col\Btree(p_k)$. Suppose
  that $t_1,\ldots,t_k$ \emph{match} $l_1,\ldots,l_k$. Then
  $\match_\cP(\vt,\vp)$ is defined.
\end{lem}

\begin{prf}
  We proceed by induction on the structures of $l_i$
  (matching the cases of the $\match_\cP$ judgement)
  
  \begin{itemize}
  \item $l_1=x_1,\ldots,l_n= x_n$. In this case, the only applicable
    case for $\th_{min}$ is the variable case. If $x_i=x_j$, then
    $t_i=t_j$. Furthermore $p_i=\al_i$ for some variable $\al_i$ and
    again, $\al_i=\al_j$ if and only if $x_i=x_j$, by linearity of
    $\al_i$ and $\al_j$ in $\G$. Therefore if $\al_i=\al_j$, then
    $t_i=t_j$, and $\match_\cP(\vt,\vp)$ is defined.
  \item $l_i=\Leaf$. In this case the only applicable rule is the leaf
    rule, and $p_i=\leaf$ and $t_i=\Leaf$. By
    induction $\match_\cP(\vt,\vp)$ is defined.
  \item $l_i=\Node\ l_i^1\ l_i^2$. In this case we apply the node
    rule, and we have $p_i=\node(p_i^1,p_i^2)$. Again, we have
    $t_i=\Node\ t_i^1\ t_i^2$, and we may conclude by the induction
    hypothesis.
  \end{itemize}
  
  \cqfd
\end{prf}

Our reason for defining pattern matching is to provide the ``closest''
possible pattern semantics for a term. In fact we have the following
result, which states that any valuation $\t$ such that $t$ is in
$\I{\Btree(\al)}_\t$ can be ``factored through'' $\match(t,p)$:

\begin{lem}\label{dp_prf:lem:exact_sem}
  Suppose that $\vt$ is a tuple of strongly normalizing terms, that
  $\vec{\al}$ is a
  tuple of pattern variables, and $\t'$ is a valuation that verifies:
  \[\all i, t_i\in\I{\Btree(\al_i)}_{\t'} \]
  Suppose in addition that $\vp$ are minimal patterns such that
  $\match_\cP(\vt,\vp)$ is defined and equal to $\t$. Let $\phi$ be
  the substitution that sends $\al_i$ to $p_i$. Then
  \[\all i, \t\circ\phi(\al_i)\subpat \t'(\al_i) \]
\end{lem}

\begin{prf}
  We proceed by induction on the judgment $\match_\cP(\vt,\vp)$.
  
  \begin{itemize}
  \item $p_i=\b_i$ for each $p_i$, and therefore
    $\phi(\al_i)=\b_i$. In that case, $\t\circ\phi(\al_i)=\{pat(v)\mid
    v\mbox{ normal form of }t_i\}$. Furthermore, $t_i\redmatch
    \t'(\al_i)$. Take some $v$ a normal form of $t_i$. We have some
    $q\in\t'(\al_i)$ such that $v\redmatch q$. We then verify that
    $pat(v)\subpat q$, which implies
    $\t\circ\phi(\al_i)\subpat\t'(\al_i)$
  \item $p_i=\leaf$. In this case, $\t'\circ\phi(\al_i)=\leaf$. By
    $t_i\redmatch \t'(\al_i)$ and $t_i=\Leaf$, we have that
    $\t'(\al_i)$ contains $\leaf$ or $\wild$, and in each case we can
    conclude.
  \item $p_i=\node(p_i^1,p_i^2)$. In this case,
    $t_i=\Node\ t_1^1\ t_i^2$, and 
    \[\t\circ\phi(\al_i)=\{\node(r_1,r_2)\mid r_1\in p_i^1\t\et
    r_1\in p_i^2\t\}\]
    By $t_i\redmatch \t'(\al_i)$ we have for each normal form $v$ of
    $t_i$ some $q$ in $\t'(\al_i)$ such that $v\redmatch q$. In
    addition $v$ is of the form $\Node\ v_1\ v_2$, where $v_1$ is a
    normal form of $t_i^1$ and $v_2$ is a normal form of $t_i^2$. From
    this we get that either $q=\wild$, in which case we are done, or
    $q=\node(q_1,q_2)$ with $v_1\redmatch q_1$ and $v_2\redmatch
    q_2$. In this case we apply the induction hypothesis to deduce
    that there is some $r_1\in p_i^1\t$ and $r_2\in p_i^2\t$ such that
    $r_1\subpat q_1$ and $r_2\subpat q_2$, and thus
    $\node(r_1,r_2)\subpat\node(q_1,q_2)$.
  \end{itemize}

  \cqfd
\end{prf}


\begin{dfn}
  We define the following order $>_{dp}$ on pairs $(f,\vt)$ with
  $f\in\S$ and $\vt$ a tuple of terms:
  \[(f,\vt)>_{dp}(g,\vu)\AA \ex \vt',n=f^\sharp(\vp)\red
  g^\sharp(\vq),\ \vt\red^* \vt'\et (n,\vt',\vu)\mbox{
    is a link}\]
  That is, if $\vt$ reduces to $\vt'$ such that there is a link
  between $\vt'$ and $\vu$, and where the associated node corresponds
  to a call from $f$ to $g$.

\end{dfn}

\begin{lem}
  If the conditions of theorem \ref{ho_dp:thm:sn} are satisfied then
  the order $>_{dp}$ is well-founded.
\end{lem}

\begin{prf}
  Any infinite decreasing sequence
  $(f_1,\vt_1)>_{}(f_1,\vt_2)>_{}\ldots$ gives rise to an infinite
  chain, which is not possible by theorem
  \ref{dp_prf:thm:finite_chains}.
\end{prf}

We have enough to prove the main theorem, that is correctness of
defined symbols.

\begin{thm}\label{dp_prf:thm:correct_def_fun}
  Suppose that the conditions of theorem \ref{ho_dp:thm:sn} are
  satisfied. Then for each $f\in\S$ and each valuation $\t$,
  $f\in\I{\tau_f}$.
\end{thm}

\begin{prf}
  Suppose that
  $\tau_f=\all\vec{\al}.\Btree(\al_1)\a\ldots\a\Btree(\al_k)\a
  T_f$. Take $\t$ a valuation and $t_1,\ldots,t_n$ in
  $\I{\Btree(\al_1)}_\t,\ldots,\I{\Btree(\al_k)}_\t$. We need to show
  that
  \[f\ t_1\ldots t_n\in\I{T_f}_\t \]
  Note that each $t_i$ is strongly normalizing. We proceed first by
  induction on $\vt$ ordered by strict reduction. As $t=f\ \vt$ is
  neutral, it suffices to consider all the one step reducts $t'$ of
  $t$. These reducts are of two forms:
  
  \begin{itemize}
  \item $t'=f\ t_1\ldots t_i'\ldots t_k$ with $t_i\red t_i'$. We
    conclude by the induction hypothesis.
  \item There is some rule $l\red r\in\cR$, and some substitution
    $\s$ such that $\erase{l}\s=t$, and $\erase{r}\s=t'$. We
    then proceed by induction on $(f,\vt)$ ordered by $>_{dp}$. We
    have by hypothesis that there is some context $\G$ and some
    derivation $\G\th_{min}l_i\col \Btree(p_i)$ for each $i$, and a
    derivation $\G\th r\col T_f\phi$, with $\phi$ the substitution
    that sends $\al_i$ to $p_i$.

    By lemma \ref{dp_prf:lem:min_typ_match}, $\match_\cP=\psi$ is
    defined. We therefore have $t_i\in\I{\Btree(p_i)}_\psi$ for each
    $i$, which gives $t_i\in\I{\Btree(\al_i)}_{\psi\circ\phi}$ by the
    substitution lemma. By lemma \ref{dp_prf:lem:exact_sem},
    $\psi\circ\phi\subpat\t$. We may then apply the positivity
    condition of $\tau_f$ using lemma \ref{dp_prf:lem:leq_subst} to
    deduce that $\I{T_f}_{\psi\circ\phi}\sle\I{T_f}_\t$. Therefore it
    suffices to show that $t'$ is in $\I{T_f}_{\psi\circ\phi}$, which
    is equal to $\I{T_f\phi}_\psi$ by the substitution lemma. By
    hypothesis, $\G\th r\col T_f\phi$, so we would like to apply the
    correctness theorem \ref{dp_prf:thm:rel_correct} to show that
    $t'=\erase{r}\s\in\I{T_f\phi}_\psi$. The correctness theorem
    itself can not be applied, as it takes as hypothesis the
    correctness of function symbols, which we are trying to prove. But
    we will proceed in the same manner, making essential use of the
    well-founded induction hypothesis.

    Let us first show by induction on the derivation of
    $\G\th_{min}l_i$ that for each $x\in\dom(\s)$,
    $\s(x)\in\I{\G(x)}_\psi$.
    
    \begin{itemize}
    \item $l_i=x$. We have $\s(x)=t_i\in\I{\Btree(\g)}_\psi$ with
      $\g= p_i$ and $\psi(\b_i)=\{pat(v)\mid v\mbox{ normal form of
    }t_i\}$
    \item $l_i=\Leaf$. We have nothing to show here.
    \item $l_i=\Node\ l^1\ l^2$. Simple application of the induction hypothesis.
    \end{itemize}
    
    Now we prove by induction on the derivation of $\G\th r\col
    T_f\phi$ that $\erase{r}\s\in\I{T_f\phi}_\psi$. We can exactly
    mimic the proof of theorem \ref{dp_prf:thm:rel_correct}, except
    for the {\bf symb} case. In this case, there is a $g$ such that
    $r=g\vq$, and if $\tau_g=\all
    \vec{\b}.\Btree(\b_1)\a\ldots\a\Btree(\b_m)\a T_g$, we need to
    show that, for some extension $\psi'$ of $psi$,
    $g\in\I{\Btree(q_1)\a\ldots\a\Btree(q_m)\a T_g}_{\psi'}$. Recall
    the induction hypothesis on $(f,\vt)$, which states that for every
    $\t$, if $(f,\vt)>_{dp}(g,\vu)$, then $g\vu\in\I{T_g}_\t$. Now
    take $\t$ to be $\psi'\circ\z$ where $\z$ is the substitution that
    sends $\b_i$ to $q_i$. It suffices to show that if for
    $i=1,\ldots,m\ u_i\in\I{\Btree(\b_i)}_{\psi'\circ\z}$, then
    $(f,\vt)>_{dp}(g,\vu)$. For this we need to show that there exists
    $n\in\cG$ such that:
    
    \begin{itemize}
    \item $n=f^\sharp(\vr)\red g^\sharp(\vs)$
    \item $\match_\cP(\vt,\vr)=\t$
    \item There is an extension $\t'$ of $\t$ such that
      \[\vu\redmatch \s\t' \]
    \end{itemize}
    We just take $n$ to be the node that corresponds to the call site
    of $g\vq$. In this case, $\vr=\vp$ and $\vs=\vq$. By definition,
    $\match_\cP(\vt,\vp)$ is defined and equal to $\psi$. Then $\psi'$
    is an extension of $\psi$ and as
    $u_i\in\I{\Btree(\b_i)}_{\psi'\circ\z}=\I{\Btree(q_i)}_{\psi'}$,
    we have $u_i\redmatch q_i\psi'$.
    
  \end{itemize}
  
  \cqfd
\end{prf}

\begin{cor}
  Every well-typed term is in the interpretation of its type, that is
  \[\all \G,t,T\ \G\th t\col T\A \erase{t}\in\I{T} \]
  Where $\I{T}$ is $\I{T}_\t$ where $\t$ is the valuation that
  sends every variable to the set $\{\wild\}$.
\end{cor}

\begin{prf}
  In fact it does not matter which $\t$ we choose: let $\t$ be any
  valuation. Given a variable $x$ and a type $T$, by lemma
  \ref{dp_prf:lem:girard_cond}, $x\in\I{T}_\t$, as $x$ is neutral and
  in normal form. Given $\G\th t\col T$, we can therefore take the
  substitution $\s$ that sends every variable $x\in\dom(\G)$ to
  itself. In that case $\s(x)\in\I{\G(x)}_\t$ by the above remark, and
  by the combination of theorem \ref{dp_prf:thm:rel_correct} and
  theorem \ref{dp_prf:thm:correct_def_fun},
  $\erase{t}\s\in\I{T}_\t$. But in this case $\erase{t}\s=\erase{t}$.

  \cqfd
\end{prf}

We obtain the statement of theorem \ref{ho_dp:thm:sn} as a corollary:
every well typed term is in the interpretation of its type, but this
interpretation only contains strongly normalizing terms by lemma
\ref{dp_prf:lem:girard_cond}.

\end{document}